\documentclass[11pt,a4paper]{article}
\usepackage{jcappub}

\usepackage{graphicx}
\usepackage{color}
\usepackage{epstopdf}
\usepackage{bm}
\usepackage{amsmath}
\usepackage[amssymb]{SIunits}
\usepackage{accents}
\usepackage{slashed}

\newcommand{\dNeff}{\Delta N_{\textrm{eff}}}
\newcommand{\LASAGNA}{\textsc{lasagna}}

\newcommand{\eref}[1]{(\ref{#1})}
\newcommand{\fref}[1]{figure~\ref{#1}}
\newcommand{\sref}[1]{section~\ref{#1}}
\newcommand{\tref}[1]{table~\ref{#1}}

\renewcommand{\P}{\mathbf{P}}
\newcommand{\V}{\mathbf{V}}

\newcommand{\p}{\mathbf{p}}
\renewcommand{\k}{\mathbf{k}}
\newcommand{\q}{\mathbf{q}}
\newcommand{\w}{\mathbf{w}}
\renewcommand{\v}{\mathbf{v}}
\newcommand{\feq}{f_{\textrm{eq}}}
\newcommand*{\pbar}[1]{\accentset{(-)}{#1}}

\begin{document}

\begin{flushright}
{\large \tt 
INT-PUB-15-025}  
\end{flushright}

\title{Active--sterile neutrino oscillations in the early Universe with full collision terms}

\author[a,b]{Steen Hannestad}

\author[a,c]{Rasmus Sloth Hansen}

\author[d]{Thomas Tram}

\author[c]{Yvonne Y.~Y.~Wong}
\affiliation[a]{Department of Physics and Astronomy,
 Aarhus University, 8000 Aarhus C, Denmark}
\affiliation[b]{Aarhus Institute of Advanced Studies,
 Aarhus University, 8000 Aarhus C, Denmark}
\affiliation[c]{School of Physics, The University of New South Wales, 
Sydney NSW 2052, Australia}
\affiliation[d]{Institute of Cosmology and Gravitation, University of Portsmouth, Portsmouth PO1 3FX, United Kingdom
}
\emailAdd{sth@phys.au.dk}
\emailAdd{rshansen@phys.au.dk}
\emailAdd{thomas.tram@port.ac.uk}
\emailAdd{yvonne.y.wong@unsw.edu.au}

\date{\today}

\abstract{
Sterile neutrinos are thermalised in the early Universe via oscillations with the active neutrinos for certain mixing parameters. The most detailed calculation of this thermalisation process involves the solution of the momentum-dependent quantum kinetic equations, which track the evolution of the neutrino phase space distributions.  Until now the collision terms in the quantum kinetic equations have always been approximated using equilibrium distributions, but this approximation has never been checked numerically. In this work we revisit the sterile neutrino thermalisation calculation using the {\it full} collision term, and compare the results with various existing approximations in the literature. 
We find a better agreement than would naively be expected, but also identify some issues with these approximations that have not been appreciated previously. These include an unphysical production of neutrinos via scattering and the importance of redistributing momentum through scattering, as well as details of Pauli blocking. Finally, we devise a new approximation scheme, which improves upon some of the shortcomings of previous schemes.
}

\maketitle


\section{Introduction}

Light sterile neutrinos with masses around an eV have long been postulated as an explanation for the anomalous results seen in a number of short-baseline neutrino oscillation experiments~(see, e.g.,~\cite{Giunti:2013aea} for a review).  Although intriguingly each experiment appears to point to a slightly different mass value, the preferred ranges of masses and active--sterile mixing angles are such that, within standard cosmology, the sterile neutrino state inevitably becomes thermalised in the early Universe.  The presence of an additional thermalised neutrino species in the eV-mass range is however in conflict with  observations of  the cosmic microwave background (CMB) anisotropies and the large-scale structure (LSS) distribution: the best limit from the ESA Planck mission and other astrophysical observations on the number of thermalised neutrino species currently  stands at 
$N_{\rm eff} = 3.04 \pm 0.18$ (68\% C.I.)~\cite{Planck:2015xua}. 
Thus, if the short-baseline anomalies are to be interpreted in terms of active--sterile neutrino oscillations despite their mutual tension, some mechanism to reconcile the sterile state with cosmology would be required.

Several such mechanisms to circumvent the CMB/LSS limits on sterile neutrinos have been investigated in the past, some of which employ a large lepton asymmetry~\cite{Foot:1995bm,Hannestad:2012ky,Mirizzi:2012we,Saviano:2013ktj, Chu:2006ua} or interactions among the sterile neutrinos~\cite{Hannestad:2013ana,Dasgupta:2013zpn,Saviano:2014esa,Mirizzi:2014ama,Chu:2015ipa} to suppress the production of sterile neutrinos in the early Universe. What these two mechanisms have in common is that they both delay the production of sterile neutrinos until at the earliest the neutrino decoupling epoch~($T \sim {\rm MeV}$), after which the active neutrino states cannot be fully repopulated even if the active--sterile oscillation probability should become large.
This limits the total number of neutrinos populating the Universe, and hence avoids an unacceptably large energy density in relativistic particles.

To accurately track the sterile neutrino thermalisation process---in the presence of large lepton asymmetry, self-interactions, or otherwise---requires that we solve the quantum kinetic equations (QKEs), which describe the evolution of the density matrix of the active and sterile neutrino phase space distributions in the presence of flavour oscillations and scattering (forward and non-forward).  The formal expressions for all components of the QKEs up to order $G_F^2$ are well known~\cite{McKellar:1992ja, Sigl:1992fn}.   However, as the collision integrals describing the non-forward scattering are generally quite complicated and nonlinear,  it is customary to approximate them in numerical solutions of the QKEs.  Common approximations include neglecting Pauli blocking and feedback from the real-time neutrino phase space distributions, as well as ignoring the electron mass in the evaluation of the scattering matrix elements~\cite{Bell:1998ds, Chu:2006ua}. 

In many ways these were reasonable approximations in their time.  However, as observational precision improves, it also becomes necessary to examine more closely their precise impact on the observables.  A conservative and naive estimate of the error due to neglecting the Pauli blocking factors, for example, is $\sim 10\%$~\cite{Bell:1998ds} for active--sterile conversion before the neutrino decoupling epoch, which in view of Planck's sensitivity to $N_{\rm eff}$ already appears to be pushing the limits of validity.
As the conversion temperature approaches the decoupling epoch, the details of how the active neutrino states are repopulated and their oscillations to sterile states damped through collisions  can only have an even larger impact; apart from  thermalisation of the sterile neutrino and hence the $N_{\rm eff}$~parameter, the detailed repopulation mechanism affects in principle also the active neutrino energy spectra, which could subsequently alter the weak reaction rates during big bang nucleosynthesis~(BBN).

In this paper we include for the first time the full collision integral in the solution of the QKEs for active--sterile neutrino oscillations.  
For simplicity we restrict our attention to a two-neutrino model without lepton asymmetries or sterile neutrino self-interaction.
We  focus on oscillations between the electron and the sterile neutrinos described by a mass squared difference~$\delta m^2$ and vacuum mixing angle~$\theta$, 
although we will keep the equations general and bear in mind also the cases of sterile neutrino oscillations with muon and tau neutrinos. 
We assume  the sterile neutrino to be heavier than the active neutrinos as this is the most interesting case for mass squared differences above $\delta m^2 \sim 0.1~\electronvolt^2$: 
current cosmological limits on the sum of the neutrino masses are in the sub-eV range~\cite{Archidiacono:2013fha,Costanzi:2014tna,Planck:2015xua}, and would be violated if the active neutrinos were heavier than the sterile state for the $\delta m^2$ values of interest.

The rest of the paper is organised as follows.  We begin with an introduction to the QKEs in  \sref{sec:equations}, where we describe the repopulation of the active neutrinos and the damping of flavour oscillations from collisions. We discuss  various approximations found in the literature, and devise new, improved approximations that incorporate more of the relevant physics. In \sref{sec:numresults} we first test our implementation of the full collision term for convergence, before proceeding to systematically compare the various approximate solutions with the full result.  We find that although the deviations are smaller than expected for most neutrino mixing parameters, there are some systematic biases that can be significantly reduced using our new approximation scheme. We give our conclusions in \sref{sec:conclusion}.  Throughout the paper we employ a unit system in which $c = \hbar = k_B = 1$, and express all dimensionful quantities in powers of eV.


\section{Quantum kinetic equations}
\label{sec:equations}

We consider oscillations between an active neutrino flavour $\nu_\alpha$, where $\alpha = e$, $\mu$ or $\tau$, and a sterile flavour $\nu_s$ in an ensemble of neutrinos 
in the early universe.   The density matrices~${\bm \rho}(k)$ encode the flavour content and coherence of the ensemble, and are conveniently expressed 
in terms of polarisation vectors $(P_0(k),\mathbf{P}(k))$, i.e.,
\begin{equation*}
  {\bm \rho}(k) =
  \begin{pmatrix}
    {\rho}_{\alpha \alpha}(k) & \rho_{\alpha s}(k) \\  \rho_{s\alpha}(k) & \rho_{ss}(k)
  \end{pmatrix}
  = \frac{1}{2}f_0(k)[P_0(k) \mathbb{1} + \mathbf{P}(k)\cdot {\bm \sigma}],
\end{equation*}
where $f_0(k)$ denotes a reference phase space density, which we take here to be the relativistic Fermi--Dirac distribution with vanishing chemical potential,%
\footnote{Henceforth we shall reserve $f_0$ to denote exclusively the relativistic Fermi--Dirac distribution with zero chemical potential, and use $f_{\rm eq}$ to indicate an equilibrium distribution of a nonspecific form.}
 and ${\bm \sigma} = \{\sigma_x, \sigma_y, \sigma_z\}$ are the Pauli matrices. In this convention the active and sterile neutrino phase space distributions are given, respectively, by
\begin{equation}
\begin{aligned}
f_{\nu_\alpha}(k) &= \frac{1}{2} [P_0 (k) + P_z(k)], \\
f_{\nu_s}(k) &= \frac{1}{2} [P_0 (k) - P_z(k)], \\
\end{aligned}
\end{equation}
and the  QKEs that govern their evolution can be written as~\cite{McKellar:1992ja,Bell:1998ds}
\begin{equation}
\begin{aligned}
  \label{eq:QKE1}
  \dot\P(k) &= \V(k) \times \P(k) + \frac{R_\alpha(k)}{f_0(k)} \hat{\mathbf{z}} - D(k) \P_T(k) ,\\
  \dot P_0(k) &= \frac{R_\alpha(k)}{f_0(k)}.
\end{aligned}
\end{equation}
Here, $\V(k) \equiv V_x(k) \hat{\mathbf{x}} + V_z(k) \hat{\mathbf{z}}$ contains the vacuum oscillation term as well as the matter potential from forward scattering, and the components are given respectively by
\begin{equation}
\begin{aligned}
  V_x(k) &= \frac{\delta m^2}{2k} \sin 2\theta,\\
  V_z(k) &= -\frac{\delta m^2}{2k} \cos 2\theta - \frac{7\pi^2G_F}{45\sqrt{2}M_Z^2} k T^4 (n_{\nu_\alpha} + n_{\bar{\nu}_\alpha}) g_\alpha,
\end{aligned}
\end{equation}
where $G_F$ is the Fermi constant, $M_Z$ the mass of the $Z$ boson, $n_i$ the number density normalised to the equilibrium value, $g_{\mu/\tau} = 1$, and $g_e = 1+4\sec^2\theta_W/(n_{\nu_e}+n_{\bar\nu_e})$ with  the Weinberg angle $\theta_W$.
Finally, $R_\alpha(k)$ and $D(k)$ are respectively the repopulation and damping terms, which we describe in more detail in section~\ref{sec:repop}.
As we assume a vanishing lepton asymmetry, the same QKEs apply to both neutrinos and antineutrinos, and $n_{\nu_\alpha} = n_{\bar \nu_\alpha}$.


\subsection{Repopulation and damping}
\label{sec:repop}

The repopulation and damping terms are integrals over the matrix elements for annihilation and elastic scattering processes. Beginning with equation~(24) of~\cite{McKellar:1992ja}, which includes Pauli blocking and appears also in~\cite{AP:1981},
we find these integrals to be 
\begin{equation}
\begin{aligned}
  R_\alpha(k) &=  2\pi \int d \Pi_{k'} d \Pi_{p'}d \Pi_{p} \; \delta_E(kp| k'p') \\
  &\hspace{1.5cm} \times \sum_i \mathcal{V}^2[\nu_\alpha(k),\bar \nu_\alpha (p)| i(k'),\bar i(p')] \left[f_i(E_{k'})f_{\bar{i}}(E_{p'})(1-f_{\nu_\alpha}(k))(1- f_{\bar\nu_{\alpha}}(p))\right. \\
  &\hspace{5cm} - \left.f_{\nu_\alpha}(k) f_{\bar\nu_{\alpha}}(p)(1-f_i(E_{k'}))(1-f_{\bar{i}}(E_{p'}))\right] \\
  &\hspace{1.9cm} + \sum_j\mathcal{V}^2[\nu_\alpha(k),j (p)| \nu_\alpha(k'), j(p')] \left[f_{\nu_\alpha}(k')f_j(E_{p'})(1-f_{\nu_\alpha}(k)) (1-f_j(E_p))\right.\\
  &\hspace{5cm}\left.- f_{\nu_\alpha}(k) f_j(E_p)(1-f_{\nu_\alpha}(k'))(1-f_j(E_{p'}))\right],\label{eq:Ralpha}
  \end{aligned}
  \end{equation}
  and
 \begin{equation}
 \begin{aligned}
  D(k) &= \pi \int d \Pi_{k'} d \Pi_{p'} d \Pi_{p} \; \delta_E(kp|k'p') \\
  &\hspace{1.5cm} \times \sum_i \mathcal{V}^2[\nu_\alpha(k),\bar\nu_\alpha(p)| i(k'),\bar i(p')] \left[f_i(E_{k'})f_{\bar i}(E_{p'}) (1-f_{\bar \nu_\alpha}(p)) \right. \\
  &\hspace{5cm} \left.+ f_{\bar \nu_\alpha}(p) (1-f_{\bar i}(E_{p'})) (1-f_{i}(E_{k'})) \right] \\
  &\hspace{1.9cm} + \sum_j \mathcal{V}^2[\nu_\alpha(k),j(p)|\nu_\alpha(k'),j(p')] \left[f_{\nu_\alpha}(k')f_j(E_{p'}) (1-f_j(E_p))\right.  \\
  &\hspace{5cm} \left.+ f_j(E_p) (1-f_j(E_{p'})) (1-f_{\nu_\alpha}(k'))\right], \label{eq:Deq}
\end{aligned}
\end{equation}
where we have used the shorthand $d \Pi_p  \equiv d^3\p/(2\pi)^3$, $\delta_E(k p | k' p')  \equiv \delta^{(1)} (E_k+E_p-E_{k'}-E_{p'})$ is the 1D Dirac delta function,  the summation index $i$ runs over all spectator neutrino flavours (i.e., $\nu_\beta$ where $\beta \neq \alpha$) and the electron, while $j$  runs in addition over all their antiparticles as well as the oscillating neutrino and antineutrino. The terms $\mathcal{V}^2$ are
\begin{equation}
  \mathcal{V}^2[a(p),b(k)| c(p'),d(k')] = (2\pi)^3 \delta^{(3)}(k+p,k'+p') N^2_aN^2_bN^2_cN^2_d S|M|^2(a(p),b(k)| c(p'),d(k')),
\end{equation}
where $N_i = \sqrt{1/2E_i}$,%
\footnote{The prefactor $N_i = \sqrt{1/2 E_i}$  used here differs from the definition in~\cite{McKellar:1992ja} due to the normalisation of the Dirac spinor.  Our choice of normalisation gives the completeness relations $u(p_i) \bar{u}(p_i) = \slashed{p_i}+m_i$ and $v(p_i) \bar{v}(p_i) = \slashed{p_i}-m_i$. }
 $E_i$  denotes the energy of particle $i$, and $S|M|^2(a(p),b(k)| c(p'),d(k'))$ is the squared matrix element for the  forward process $a(p)b(k) \rightarrow c(p') d(k')$, summed (but not averaged) over initial and final spins, and 
 symmetrised over identical particles in the initial and the final state.   If two $\nu_\alpha$s are present in the initial state, then $S|M|^2$ must additionally be multiplied by 2 to account for the fact that  $\nu_\alpha(k) \nu_\alpha (p) \to \ldots$ and $\nu_\alpha (p) \nu_\alpha(k)\to \ldots $ constitute two identical processes.

The first part of both the repopulation and the damping integrals~(\ref{eq:Ralpha}) and~(\ref{eq:Deq})
pertains to annihilation processes, while the rest describes scattering processes. 
The repopulation integral~(\ref{eq:Ralpha}) incorporates Pauli blocking in the form of additional multiplicative factors of the form $(1-f_j)$ for every particle~$j$ in the final state of both the forward and reverse processes, 
and conforms with expectations.    For the damping integral~(\ref{eq:Deq}), however, Pauli blocking enters in a way that may not be entirely intuitive.  Compared to the expression used by McKellar and Thomson~\cite{McKellar:1992ja}, 
\begin{equation}
 \begin{aligned}
  D(k) = &\pi \int d \Pi_{k'} d \Pi_{p'} d \Pi_p \; \delta_E(kp|k'p') \\
  & \times \sum_i \mathcal{V}^2[\nu_\alpha(k),\bar\nu_\alpha(p)| i(k'),\bar i(p')] f_{\bar \nu_\alpha}(p) + \sum_j \mathcal{V}^2[\nu_\alpha(k),j(p)|\nu_\alpha(k'),j(p')]  f_j(E_p),
  \end{aligned}
\end{equation}
we find two modifications for each interaction process: one additional term (the ``first term'') and new multiplicative factors in the second term.
In terms of the evolution of the density matrix ${\bm \rho}$, Pauli blocking enters  the equation of motion as a multiplicative matrix factor of the form $\delta_{ij} - \rho_{ij}$ (see equation~(24) of~\cite{McKellar:1992ja}).  The appearance of the first term can be traced to the off-diagonal part of this matrix factor, while the second term includes a factor from the matrix diagonal. 
Because the two corrections differ in sign, they cancel one another to some extent.

A naive introduction of Pauli blocking into the damping integral~(\ref{eq:Deq}) might lead one to miss the first term, which would result in an underestimation of the damping.   However, as it turns out, the negative correction contained in the second term dominates anyhow when using equilibrium distributions,  so that the effect of Pauli blocking is similar to what would naively be expected, albeit smaller.

In appendix~\ref{sec:full} we evaluate the repopulation and damping integrals~(\ref{eq:Ralpha}) and~(\ref{eq:Deq}) using the technique described by Hahn-Woernle, Pl\"umacher and Wong~\cite{HahnWoernle:2009qn}. With this approach we can reduce the nine-dimensional integral to three dimensions. Of these it is possible to perform one integral analytically, but the remaining two must be calculated numerically. 


\subsection{Approximation schemes}
\label{sec:approx_intro}

The customary way to treat the collision terms is to assume most of the particles to be in equilibrium with the background photons.   This simplifies the integrals so that solving the final expression is less numerically demanding.
Assuming that Pauli blocking can be neglected and that all species follow equilibrium distributions except for the one being repopulated,  the repopulation and damping terms evaluate in what we shall call the equilibrium approximation~\cite{Bell:1998ds,Kainulainen:2001cb,Hannestad:2012ky} to
\begin{align}
  R_{\textrm{eq}}(k) &= \Gamma \left( f_0 - \frac{f_0}{2}\left(P_0 + P_z\right)\right),   \label{eq:approxR}\\
  D_{\textrm{eq}}(k) &= \frac{1}{2}\Gamma, \label{eq:approxD}
\end{align}
where $\Gamma = C_\alpha G_F^2 x T^5$, with $x = k/T$, $C_e \approx 1.27$, and $C_{\mu,\tau} \approx 0.92$~\cite{Enqvist:1991qj,Cline:1991zb,Kainulainen:2001cb}.  While this expression makes intuitive sense in terms of 
bringing the distribution towards  equilibrium and coincides with the relaxation time approximation commonly found in the Boltzmann transport literature, it can also be derived from a firmer basis~\cite{McKellar:1992ja,Bell:1998ds}.

An alternative approximation scheme is that introduced by Chu and Cirelli~\cite{Chu:2006ua} (CC approximation), which is based on a second quantised approach~\cite{Sigl:1992fn}.
 Here,  the combined collision (damping and repopulation) term is~\cite{Chu:2006ua, Saviano:2013ktj}
\begin{equation}
  \label{eq:CC}
  \dot {\bm \rho}_{\textrm{coll}}(k) = - \frac{G_F^2 T^5}{2}  \frac{k}{\left<k\right>}\left(\left\{{\bm G}_s^2,{\bm \rho}-{\bm \rho}^{\textrm{0}}\right\} - 2 {\bm G}_s ({\bm \rho}-{\bm \rho}^{\textrm{0}}) {\bm G}_s + \left\{{\bm G}_a^2,{\bm \rho}-{\bm \rho}^{\textrm{0}}\right\} + 2 {\bm G}_a({\bm \rho}-{\bm \rho}^{\textrm{0}}) {\bm G}_a\right),
\end{equation}
where ${\bm G}_{s,a} = {\rm diag} \left( \gamma_{s,a}^\alpha,0 \right)$, with $(\gamma_s^e)^2 = 3.06$ and $(\gamma_s^{\mu,\tau})^2 = 2.22$ for scattering processes, 
and $(\gamma_a^e)^2 = 0.50$ and $(\gamma_a^{\mu,\tau})^2 = 0.28$ for annihilations~\cite{Chu:2006ua}, $\langle  k \rangle \approx 3.15 \  T$ is the average momentum, and ${\bm \rho}^0 \equiv f_0  \ \mathbb{1}$.
The CC approximation was originally applied in~\cite{Chu:2006ua} to the momentum-averaged quantum rate equations, and
to adapt it for the momentum-dependent quantum kinetic equations we have had to scale equation~(\ref{eq:CC}) by a factor  $k / \left<k\right>$~\cite{Saviano:2013ktj} to approximate the momentum dependence.
Evaluating the matrix products, we find
\begin{equation}
  \label{eq:CCmatrix}
  \dot {\bm \rho}_{\textrm{coll}}(k) = - \frac{1}{2} G_F^2 T^5 \frac{k}{\left<k\right>} 
  \begin{pmatrix}
    4 (\gamma_a^{\alpha})^2 (\rho_{\alpha \alpha} - \rho^{\textrm{0}}_{\alpha\alpha}) & \left[(\gamma_s^{\alpha})^2 + (\gamma_a^{\alpha})^2 \right] \rho_{\alpha s}\\
    \left[(\gamma_s^{\alpha 2})^2 + (\gamma_a^{\alpha})^2 \right] \rho_{s \alpha} & 0
  \end{pmatrix},
\end{equation}
which can be further recast as expressions for repopulation and damping,
\begin{align}
  \label{eq:CCR}
  R_{\textrm{CC}}(k) &= 2 G_F^2 T^5 \frac{x}{3.15} (\gamma_a^\alpha)^2\left(f_0 - \frac{f_0}{2}(P_0+P_z)\right),\\
  \label{eq:CCD}
  D_{\textrm{CC}}(k) &= \frac{1}{2} G_F^2 T^5 \frac{x}{3.15} \left[ (\gamma_s^{\alpha 2})^2 + (\gamma_a^{\alpha})^2\right],
\end{align}
compatible with equation~\eref{eq:QKE1}.


\subsubsection{Repopulation} 
\label{sec:repopulation}

Observe that in the CC approximation only the annihilation processes contribute to the repopulation term~(\ref{eq:CCR}). This approximation is reasonable for the momentum-averaged quantum rate equations where the expression was first used, but is not strictly correct in the momentum-dependent quantum kinetic equations where elastic scattering processes is also expected to contribute to the equilibration of individual momentum bins.  The equilibrium approximation~(\ref{eq:approxR}) on the other hand does take into account scattering processes.  However, in doing so, it  also sacrifices the principle of detailed balance in that scattering processes can only redistribute momentum but not repopulate a momentum bin without reference to the population of other bins.
A better solution would be to separate the two contributions, but this would require a more advanced treatment than a simple relaxation towards one equilibrium distribution.

Here, we develop a new repopulation approximation scheme that keeps the annihilation and scattering terms separate.  We call this the A/S approximation.
Looking first at annihilations, the full expression for $\nu_\alpha \bar{\nu}_\alpha$ annihilation into $a$ and $\bar a$ is an integral over $(f_a f_{\bar a} - f_{\nu_\alpha} f_{\bar \nu_\alpha})$, where we have ignored Pauli blocking factors. In the equilibrium approximation, it is assumed that $f_i \approx f_0$ for all particles $i=a,\bar{a},\bar{\nu}_\alpha$ but $\nu_\alpha$. For $a$ and $\bar a$ this is a good assumption. For~$\bar \nu_\alpha$, however, we know that it overestimates $f_{\bar \nu_\alpha}$ if $n_{\bar \nu_\alpha} < 1$. Thus, to compensate for the depletion of $\bar \nu_\alpha$, we adopt $f_{\bar{\nu}_\alpha} \approx n_{\nu_\alpha} f_0$ (remembering that $n_{\bar \nu_\alpha} = n_{\nu_\alpha}$) in the annihilation part of the A/S approximation, while keeping  $f_i = f_0$ for  $i=a,\bar{a}$.

For the scattering processes on the other hand, we must take care to preserve the neutrino number density at all times.  We accomplish this in two ways, depending on whether  the scattering process involves energy and momentum exchange between the $\nu_\alpha$ population and an external bath.
\begin{enumerate}

\item {\it Nonzero energy and momentum exchange}.  Scattering processes involving electrons, positrons and spectator neutrinos~$\pbar{\nu}_\beta$,  all of which are assumed to be in equilibrium with the photons, serve to drive the $\nu_\alpha$ population to a thermal distribution with  temperature equal to the photon temperature~$T$.
We therefore model the process using a repopulation term of the relaxation form~(\ref{eq:approxR}), but with the equilibrium distribution to which $f_{\nu_\alpha}$ relaxes
replaced by
\begin{equation}
  \label{eq:feqs}
  f_{\textrm{eq, scat}} = \frac{1}{e^{x - \xi}+1},
\end{equation}
where  $\mu = \xi T$ is a pseudo-chemical potential.\footnote{A pseudo-chemical potential appears with the same sign for both particles and anti-particles, whereas the normal chemical potential has opposite signs for particles and anti-particles.}  The $\xi$~parameter can be determined in practice from the condition that the scattering contribution to $\dot{n}_{\nu_\alpha}$ must be zero, or equivalently, that the third (not the second) kinetic moments of $f_{\textrm{eq, scat}}$ and $f_{\nu_\alpha}$ must be equal (because the collision rate is proportional to momentum).  

\item {\it Vanishing energy and momentum exchange.} Scattering amongst $\nu_\alpha$ and $\bar{\nu}_\alpha$ conserves both the energy and number densities of the active oscillating neutrinos.  Energy conservation implies that the $\nu_\alpha$ population could relax to a thermal distribution with a temperature~$T_{\nu_\alpha}$ different from the photon temperature~$T$, i.e.,
\begin{equation}
  \label{eq:feqnu}
  f_{\textrm{eq, }\nu_\alpha} = \frac{1}{e^{k/T_{\nu_\alpha} - \mu_{\nu_\alpha}/T_{\nu_\alpha}}+1} .
\end{equation}
Here, $\mu_{\nu_\alpha}$ and $T_{\nu_\alpha}$ are determined from the combined condition that the scattering contributions to $\dot n_{\nu_\alpha}$ and  $\dot N_{\nu_\alpha}$ must both be zero, where $N_{\nu_\alpha}$ is the energy density normalised to the equilibrium value.  The former constraint is equivalent to
 requiring equality between the third kinetic moments of $f_{\textrm{eq, } \nu_\alpha}$ and $f_{\nu_\alpha}$, while the latter constraint calls for equality between the fourth moments.

\end{enumerate}

Thus, the full repopulation term in the A/S approximation can now be expressed as
\begin{equation}
  \label{eq:Ras}
  \begin{aligned}
  R_{\textrm{A/S}} (k) = G_F^2 x T^5 \left\{C_{\alpha,a} \left(f_0 - \frac{n_{\nu_\alpha}f_0}{2} (P_0 + P_z)\right)\right. &+ C_{\alpha,s} \left(f_{\textrm{eq, scat}} - \frac{f_0}{2} (P_0 + P_z)\right)\\
  & \left.+ C_{\alpha,\nu} n_{\nu_\alpha}\left(f_{{\rm eq},\nu_\alpha} - \frac{f_0}{2} (P_0 + P_z)\right)\right\} ,
  \end{aligned}
\end{equation}
where $C_{e,a} = 0.180$, $C_{e,s} = 0.718$, $C_{e,\nu} = 0.407$, $C_{\mu/\tau, a} = 0.102$, $C_{\mu/\tau,s} = 0.407$, and $C_{\mu/\tau,\nu} = 0.407$.  Full expressions for the coefficients can be found in appendix~\ref{sec:dampingcoeff}.
As it turns out, the separation of the scattering processes into vanishing and non-vanishing energy exchange with an external bath is not crucial for the accuracy of the approximation;  it has been included here  mainly for consistency. If it were to be abandoned, one could simply truncate equation~\eref{eq:Ras} at the second term, and obtain a new coefficient for the scattering term by adding together the numerical values of $C_{\alpha,s}$ and $C_{\alpha, \nu}$ given above.

Finally, note that equation~(\ref{eq:Ras}) does not accommodate a large lepton asymmetry; if one were present, it would be necessary to reexamine the assumptions behind the equilibrium approximation~(\ref{eq:approxR}) and all subsequent approximations that lead to~(\ref{eq:Ras}).


\subsubsection{Damping}
\label{sssec:damping}

The damping term in equation~\eref{eq:Deq} is affected  by Pauli blocking in two ways as already discussed in section~\ref{sec:repop}. As the negative correction dominates when considering equilibrium distributions, $D_{\rm eq}(k)$ in the equilibrium approximation~\eref{eq:approxD} tends to overestimate the amount of damping.  Conversely,  Chu and Cirelli~\cite{Chu:2006ua} included only the diagonal Pauli blocking terms when calculating the numerical coefficients for  $D_{\rm CC}(k)$ in equation~\eref{eq:CCD}, 
thereby underestimating the damping.%
\footnote{The collision terms of the CC approximation have been presented in~\cite{Chu:2006ua} in their integrated form, without details of how exactly they have been computed.  However, reverse engineering suggests that they arise from  the integral $\pi \int d \Pi_k d \Pi_{k'}d \Pi_{p'}d \Pi_p \; \delta_E(kp|k'p') \sum_i \mathcal{V}^2[\nu_\alpha(k),\bar\nu_\alpha(p)| i(k'),\bar i(p')] f_{\text{eq}}(p) f_{\text{eq}}(k) (1-f_{\text{eq}}(p')) (1-f_{\text{eq}}(k'))$ for the annihilations, and $\pi \int d \Pi_k d \Pi_{k'}d \Pi_{p'}d \Pi_p \; \delta_E(kp|k'p')\sum_j \mathcal{V}^2[\nu_\alpha(k),j(p)|\nu_\alpha(k'),j(p')] f_{\text{eq}}(p) f_{\text{eq}}(k) (1-f_{\text{eq}}(p')) (1-f_{\text{eq}}(k')) $ for the scatterings, both normalised by $\int d \Pi_k f_0(k)$.}
 
Here, we propose to evaluate the damping integral~\eref{eq:Deq} again with the approximations $f_i \approx f_0$ for $i \neq \nu_\alpha, \bar{\nu}_\alpha$ and 
$f_{\nu_\alpha} = f_{\bar\nu_\alpha} \approx n_{\nu_\alpha}f_0$.  Due to Pauli blocking, these approximations lead to a damping term that is quadratic in $n_{\nu_\alpha}$:
\begin{equation}
  \label{eq:Das}
  D_{\textrm{A/S}} = \frac{1}{2} G_F^2 x T^5 \left(n_{\nu_\alpha}^2 C_{\alpha,2} + n_{\nu_\alpha} C_{\alpha,1} + C_{\alpha,0} \right),
\end{equation}
where $C_{e,2} = -0.020$, $C_{e,1} = 0.569$, $C_{e,0} = 0.692$, $C_{\mu/\tau, 2} = -0.020$, $C_{\mu/\tau, 1} = 0.499$, and $C_{\mu/\tau, 0} = 0.392$,  the negative  $C_{\alpha,2}$ values reflecting the origin of the $n_{\nu_\alpha}^2$ term in the negative part of the Pauli blocking factors (expressions for the coefficients are given in appendix~\ref{sec:dampingcoeff}).
For convenience we shall continue to call this the A/S approximation, although, unlike in the repopulation treatment, there is no strict separation between annihilation and scattering; the coefficient $C_{\alpha,1}$ is predominantly from annihilation, while $C_{\alpha,0}$ comes mainly from scattering.

Lastly, we note that it is also possible to capture some of the $k$-dependence of the coefficients by omitting the $k$-integral in the computation of the damping coefficients (see appendix~\ref{sec:dampingcoeff}), and instead fit the result with a function of the form
\begin{equation}
  \label{eq:fittingD}
  C_{\alpha,\textrm{fit}} = a + b e^{-c x} - \frac{d}{x+g},
\end{equation}
where $a$, $b$, $c$, $d$, and $g$ are constants determined by the fit. This kind of expression improves the agreement between the approximation and the full treatment 
insofar as it reproduces the sterile neutrino spectrum with greater success than the constant coefficients. It does not  however lead to significant changes in $\dNeff$ and in the active spectrum which are often the most interesting quantities.  For this reason we do not incorporate $C_{\alpha, {\rm fit}}$ in the A/S approximation.


\subsection{Numerical implementation}
\label{sec:numimp}

We employ a modified version of the public code \LASAGNA{}\footnote{Available at \url{https://github.com/ThomasTram/lasagna_public}.}~\cite{Hannestad:2013pha} to solve the QKEs assuming first the full collision term~(\ref{eq:Ralpha}) and~(\ref{eq:Deq}), and then  in the various approximation schemes discussed above.   The QKEs are solved on a fixed momentum grid, with the explicit requirement that $\dot{\P}-\dot{\bar\P}=0$ so as to enforce a zero lepton asymmetry.

\subsubsection{Approximation schemes}

Implementation of the  approximate collision terms is straightforward in the equilibrium and the CC approximation.  In the A/S approximation, however, additional root-finding routines are required to evaluate the chemical potential~$\mu$ of the normal scattering term~(\ref{eq:feqs}), as well as the temperature $T_{\nu_\alpha}$ and chemical potential $\mu_{\nu_\alpha}$ of the $\nu_\alpha \nu_\alpha$-scattering term~(\ref{eq:feqnu}).

The chemical potential of the normal scattering term satisfies the equation
\begin{align*}
\int d \Pi_k \: k \frac{1}{e^{k/T - \mu/T}+1}= \frac{1}{2} \int d \Pi_k \: k  \; f_0 (P_0 + P_z),
\end{align*}
which can be solved numerically. 
The parameter space of the $\nu_\alpha \nu_\alpha$-scattering term, on the other hand, is two-dimensional and subject to the conditions
\begin{align}
\int d \Pi_k \: k \; \frac{1}{e^{k/T_{\nu_\alpha} - \mu_{\nu_\alpha}/T_{\nu_\alpha}}+1} = \frac{1}{2} \int d \Pi_k \: k \; f_0(P_0 + P_z) , \label{eq:thirdmoment} \\
\int d \Pi_k \: k^2 \; \frac{1}{e^{k/T_{\nu_\alpha} - \mu_{\nu_\alpha}/T_{\nu_\alpha}}+1}= \frac{1}{2} \int d \Pi_k \: k^2   f_0 (P_0 + P_z). \label{eq:fourthmoment}
\end{align}
In order to solve for $\mu_{\nu_\alpha}$ and $T_{\nu_\alpha}$, we first reduce the problem to one dimension by constructing the ratio
\begin{align}
  \frac{\left(\int d \Pi_k \:k^{2} f_0 (P_0+P_z)\right)^{4/5}}
           {\int d \Pi_k \:k f_0 (P_0+P_z)}
&= \frac{\left(\int d \Pi_k \:k^{2} /(e^{k/T_{\nu_\alpha}-\mu_{\nu_\alpha}/T_{\nu_\alpha}}+1)\right)^{4/5}}{\int d \Pi_k \:k /(e^{k/T_{\nu_\alpha}-\mu_{\nu_\alpha}/T_{\nu_\alpha}}+1)}, \nonumber \\
&= \left( 2\pi^2 \right)^{1/5} \frac{\left[-24\text{Li}_5\left(-e^\frac{\mu_{\nu_\alpha}}{T_{\nu_\alpha}} \right) \right]^{4/5}}{-6 \text{Li}_4\left(-e^\frac{\mu_{\nu_\alpha}}{T_{\nu_\alpha}} \right)},   \label{eq:K}
\end{align}
using equation~\eref{eq:thirdmoment} and~\eref{eq:fourthmoment}. Here $\text{Li}_s(z)$ denotes the polylogarithm. Since the RHS depends on $\mu_{\nu_\alpha}/T_{\nu_\alpha}$ but not directly on $T_{\nu_\alpha}$, equation~(\ref{eq:K}) can be solved immediately for $\mu_{\nu_\alpha}/T_{\nu_\alpha}$ using a one-dimensional root-finding algorithm.

Once we have established $\mu_{\nu_\alpha}/T_{\nu_\alpha}$, equation~\eref{eq:fourthmoment} can be evaluated explicitly for the temperature~$T_{\nu_\alpha}$. We find:
\begin{equation}
  \label{eq:Tratio}
T_{\nu_\alpha} = \left(\frac{\int d x \: x^4 f_0 (P_0 + P_z)/2}{-24\text{Li}_5\left(-e^\frac{\mu_{\nu_\alpha}}{T_{\nu_\alpha}} \right)}\right)^{1/5} T,
\end{equation}
where $T$ is the photon temperature, and again we have $x = k/T$. Finally, we  take the $\mu_{\nu_\alpha}$ and $T_{\nu_\alpha}$ values obtained from equations~(\ref{eq:K}) and (\ref{eq:Tratio}), and 
 further tune them by solving the \emph{discretised} moments equations~(\ref{eq:thirdmoment}) and~(\ref{eq:fourthmoment}) using a 2D Newton's method initialised with these numbers.
This last step  ensures that the chosen $\mu_{\nu_\alpha}$ and $T_{\nu_\alpha}$ values do indeed satisfy conservation of number and energy densities; the untuned values might not have this desired property because of the discretisation of momentum space.  In practice, however, the amount of tuning is quite small.


\subsection{Full collision term}

For the full repopulation and damping terms~(\ref{eq:Ralpha}) and~(\ref{eq:Deq}), we use the following tricks to simplify the calculations.
Consider first the repopulation integral, which we split into three parts:
\begin{align*}
R_\alpha  = R_{\alpha,e} +R_{\alpha, \beta}+R_{\alpha, \alpha},
\end{align*}
where the second subscript on the RHS labels the contributing processes.
We use equilibrium distributions $f_{\rm eq}$ for the electrons and positrons in the processes
\begin{align}
  \nu_{\alpha}(k) e^{\pm}(p) &\rightleftarrows \nu_\alpha(k') e^{\pm}(p'),  \label{eq:ereactions1}\\
  \nu_\alpha(k) \bar\nu_\alpha(p) &\rightleftarrows e^-(k') e^+(p'),  \label{eq:ereactions2}
\end{align}
 which should be a good assumption as these particles are tightly coupled via electromagnetic interactions. This assumption enables the pre-evaluation (as opposed to real-time evaluation while solving the QKEs) of one of the two energy integrals in each of equations~\eref{eq:finalRse}, (\ref{eq:finalRsp}) and (\ref{eq:finalRae}), so that the contribution of processes~\eref{eq:ereactions1} and \eref{eq:ereactions2} to $R_\alpha$ can  be expressed as
\begin{equation}
\begin{aligned}
\label{eq:Re}
  R_{\alpha,e} (k) =& (1-f(k))\left( \int dE_{k'}R_{\alpha,e,s,1}f(k') + \int dE_p R_{\alpha,e,a,1}(1-f(p))\right)\\
  &-f(k)\left( \int dE_{k'}R_{\alpha,e,s,2}(1-f(k')) + \int dE_p f(p) R_{\alpha,e,a,2}\right),
 \end{aligned}
 \end{equation} 
with
\begin{equation}
  \begin{aligned}
  \label{eq:Resi}
    R_{\alpha,e,s,1}(k,k')  &= \int dE_p (\tilde R_{\alpha,s,e^-} + \tilde R_{\alpha,s,e^+}) f_{\rm eq}(E_{p'})(1-f_{\rm eq}(E_p)),\\
  R_{\alpha,e,a,1} (k,p)&= \int dE_{k'} \tilde R_{\alpha,a,e} f_{\rm eq}(E_{k'})f_{\rm eq}(E_{p'}),\\
  R_{\alpha,e,s,2} (k,k')&= \int dE_p (\tilde R_{\alpha,s,e^-} + \tilde R_{\alpha,s,e^+}) f_{\rm eq}(E_p)(1-f_{\rm eq}(E_{p'})) ,\\
  R_{\alpha,e,a,2} (k,p) &= \int dE_{k'} \tilde R_{\alpha,a,e} (1-f_{\rm eq}(E_{k'}))(1-f_{\rm eq}(E_{p'})),
\end{aligned}
\end{equation}
where the integration kernels~$\tilde R_{\alpha,x} (k,k',p)$ encode the kinematics of the interaction processes, and  can be read off equations~\eref{eq:finalRse}, (\ref{eq:finalRsp}) and (\ref{eq:finalRae}).
We note also that $p'$ is fixed by energy and momentum conservation once a combination of $k, k', p$ has been specified.

In the limit of a massless electron, $R_{\alpha,e,s,i}$ and $R_{\alpha,e,a,i}$ of equation~(\ref{eq:Resi}) scale trivially with temperature as $\propto T^4$.
However, as the temperature drops below $\sim 1$~MeV, the massless electron approximation also becomes increasingly tenuous.  Reinstating a nonzero electron mass significantly complicates the temperature dependence of $R_{\alpha,e,s,i}$ and $R_{\alpha,e,a,i}$; we handle this by pre-evaluating equation~\eref{eq:Resi} on a grid in $(k,p, T)$ (or $(k,k',T)$), which we interpolate in real time using a 3D spline when solving the QKEs.

We use the same equilibrium approximation for the spectator neutrinos $\nu_\beta$ and antineutrinos $\bar{\nu}_\beta$ in the processes
\begin{align*}
    \nu_{\alpha}(k) \pbar{\nu}_\beta(p) &\rightleftarrows \nu_\alpha(k') \pbar{\nu}_\beta(p'),  \\
  \nu_\alpha(k) \bar\nu_\alpha(p) &\rightleftarrows \nu_\beta(k') \bar{\nu}_\beta(p'), 
\end{align*}
where $\alpha \neq \beta$, and $\pbar{\nu}_\beta$ are assumed to be non-oscillating. Thus $R_{\alpha, \beta}$ and the associated kernels $R_{\alpha, \beta, s,i}$ and $R_{\alpha, \beta, a,i}$ are, save for the $e \to \beta$ relabelling,  formally given by equations~\eref{eq:Re} and~\eref{eq:Resi} respectively, and the corresponding expressions for $\tilde R_{\alpha,x} (k,k',p)$ can be read off equations~\eref{eq:finalRsbeta}, \eref{eq:finalRsbetabar} and \eref{eq:finalRabeta}.
Since the massless neutrino limit always holds in our considerations, the temperature dependence of $R_{\alpha, \beta, s,i}$ and $R_{\alpha, \beta, a,i}$ is trivial, and the integrals need only be pre-evaluated on one 2D $(k,p)$ (or $(k,k')$) grid.

Finally, for those processes involving only the active oscillating neutrinos and/or antineutrinos, i.e.,
\begin{align*}
\nu_{\alpha}(k) \pbar{\nu}_\alpha(p) &\rightleftarrows \nu_\alpha(k') \pbar{\nu}_\alpha(p'), 
\end{align*}
we evaluate the full double integrals~\eref{eq:finalRsalpha} and~\eref{eq:finalRsalphabar} without approximation, since the momentum distributions of $\nu_\alpha$ and $\bar{\nu}_\alpha$ are {\it a priori} unknown quantities. 

For the damping term, we see from equations~\eref{eq:Ralpha} and~\eref{eq:Deq} that it differs from the repopulation term only in the missing factors of $f(k)$ and $1-f(k)$.
This means that the same simplifications of the repopulation integral discussed above and consequently all of the pre-evaluated integrals apply also to the damping term.
For example, contribution of processes~\eref{eq:ereactions1} and \eref{eq:ereactions2} to $D_\alpha$ can  be expressed as 
\begin{equation}
\begin{aligned}
\label{eq:De}
  D_{\alpha,e} (k) =& \left( \int dE_{k'}R_{\alpha,e,s,1}f(k') + \int dE_p R_{\alpha,e,a,1}(1-f(p))\right)\\
  &-\left( \int dE_{k'}R_{\alpha,e,s,2}(1-f(k')) + \int dE_p f(p) R_{\alpha,e,a,2}\right),
 \end{aligned}
 \end{equation} 
where $R_{\alpha,e,s,i}$ and $R_{\alpha,e,a,i}$ are exactly the pre-evaluated quantities of equation~(\ref{eq:Resi}).


\section{Numerical Results}
\label{sec:numresults}

The main quantity we wish to study is the change in the energy density of neutrinos due to a sterile neutrino population,
\begin{equation}
  \label{eq:Neff}
  \dNeff = N_{\rm{eff, \,} \alpha} + N_{\rm{eff, \,} s} - 1 = \frac{120}{7\pi^4 T^4} \int dk (f_{\nu_\alpha}(k) + f_{\nu_s}(k))k^3 - 1.
\end{equation}
This one quantity affects directly the Hubble expansion rate, making it possible to constrain $\dNeff$ using various cosmological observations.


\subsection{Numerical convergence}
\label{sec:convergence}

An important concern in the solution of the QKEs is detailed balance.  If detailed balance is not fulfilled, at least so to a very good approximation, it will lead to unphysical excitation of the sterile neutrinos.  As discussed in section~\ref{sec:repopulation}, detailed balance depends on the implementation of and approximations applied to the repopulation term; it is violated, for example, by the equilibrium approximation already at the level of the equation.

\begin{figure}
\center
\includegraphics[width=0.7\columnwidth]{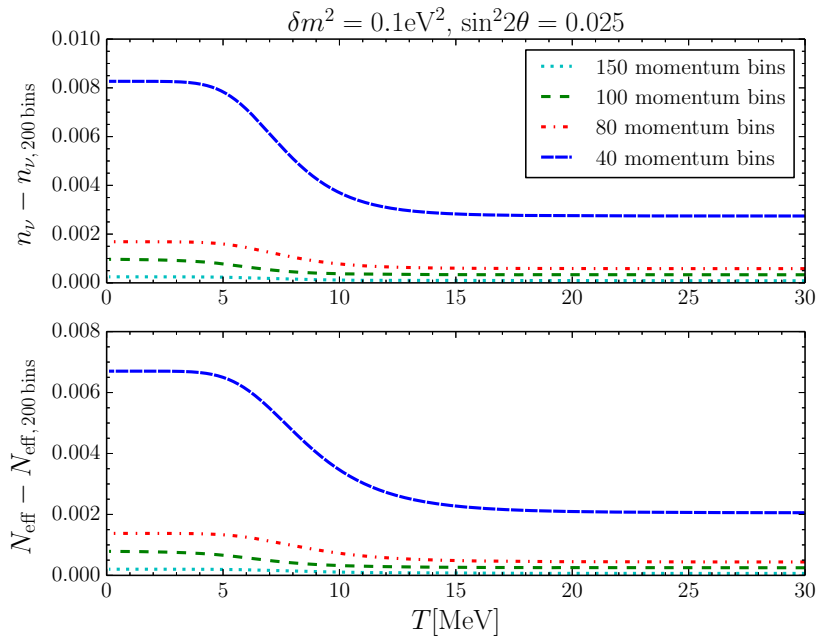}
\caption{Convergence  test using the full collision term. We compare the difference in the neutrino number and energy densities between using 200 momentum bins and 40 (blue long dashed), 80 (red dot-dashed), 100 (green dashed) and 150 bins (cyan dotted).\label{fig:convergence}}
\end{figure}

The full repopulation term, even in the presence of simplifications introduced in section~\ref{sec:numimp}, preserves detailed balance in principle. In practice, however, numerical solution  of the QKEs using the full collision term requires that we sum over a set of discretised momentum bins at every time step.  This discretisation can potentially violate detailed balance, unless the number of momentum bins is sufficiently large. In \fref{fig:convergence}, we  solve the QKEs for  the normalised  neutrino number density $n_\nu$ and effective energy density $N_{\rm eff}$ using the benchmark mixing parameter values
$\delta m^2 = 0.1~{\rm eV}^2$ and  $\sin^2 2 \theta = 0.025$, employing variously  40, 80, 100 and 150 bins; the outcomes are expressed relative to the 200 bin results.  Beyond 80 bins, convergence towards the 200 bin results to better than 0.002 across the whole temperature range of interest is immediately manifest.

Comparing the different bin choices, the offset at high temperatures originates simply in the improved representation of the distribution function as we increase the number of bins. The additional discrepancy at  $T \lesssim10$~eV is likely to have arisen from a very small deviation from detailed balance, since this is temperature regime at which the thermalisation process is most efficient.  For the remainder of the analysis we use 100 bins which gives an absolute error of $\sim 0.001$ for the benchmark mixing parameter values.
This choice is a compromise between accuracy and speed, as the evaluation of the collision integrals scales with the number of momentum bins cubed:  higher resolutions rapidly become prohibitively expensive in terms of CPU time.

\begin{figure}%
\center
\includegraphics[width=0.7\columnwidth]{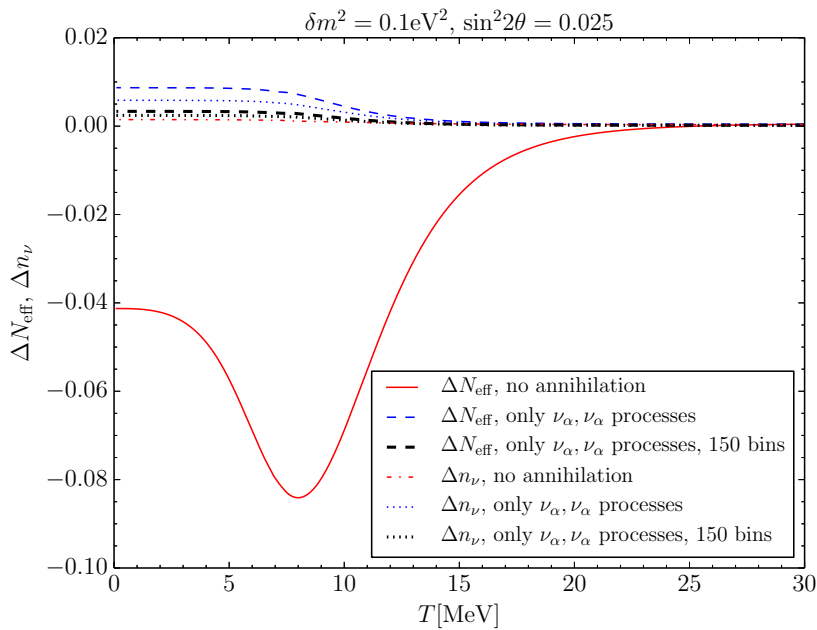}%
\caption{Different contributions to the collision terms and their effects on the neutrino number and energy densities relative to the no-oscillations case.  The red solid and dot-dash lines denote a scenario with only scattering and no annihilation.  The blue dashed and dotted lines represent scattering only amongst the oscillating active neutrinos $\nu_\alpha$; the thick black dashed and dotted lines denote the same scenario, but now computed using 150 momentum bins instead of the canonical 100.}%
\label{fig:conserve}%
\end{figure}

The full collision term sources from a variety of scattering and annihilation processes of the oscillating active neutrinos with electrons, positrons, spectator active neutrinos, and the oscillating active neutrinos themselves. Since we have computed their individual contributions explicitly, we can now also study their individual effects on the sterile neutrino thermalisation process.
This is a sanity check, but also serves to gauge the level to which our implementation of repopulation conserves energy and number densities  and hence fulfils detailed balance.
To this end, we consider in \fref{fig:conserve} two scenarios without annihilation, and compute the corresponding changes in the neutrino energy and number densities relative to the no-oscillation case for the benchmark mixing parameter values.

In the first scenario, we include all scattering processes (red solid and dot-dash lines)  and find that the neutrino number density is conserved to an excellent degree, while the energy density drops below that of the no-oscillation case. This decrease can be understood as follows.  Energy is removed from the oscillating active neutrino ${\nu}_\alpha$ population through conversion to sterile neutrinos beginning at the low end of the momentum distribution.  Some of these low-momentum active states are refilled from higher-momentum states  through ${\nu}_\alpha$ scattering with electrons, positrons, and the spectator neutrinos.  This effectively reduces the total energy contained in the combined active and sterile neutrino  population, where the surplus energy has been  absorbed into the background of $e^+$, $e^-$, and $\pbar{\nu}_\beta$.

The second scenario (blue dashed and dotted lines) consists of only scattering processes amongst the active oscillating neutrinos themselves.  Number density  conservation is again satisfied to a good degree.  Energy density is likewise approximately conserved; this is expected, as the isolated $\nu_\alpha$ population has no contact with the background plasma with which to exchange energy.   Observe that the degree of non-conservation is a larger here than in the first scenario.  This is because the evaluation of the $R_{\alpha,\alpha}$ collision integrals~\eref{eq:finalRsalpha} and~\eref{eq:finalRsalphabar}  for the $\nu_\alpha \nu_\alpha$ processes are more sensitive to numerical inaccuracies owing to their nonlinear dependence on the distribution function $f_{\bar{\nu}_\alpha}$ (see section~\ref{sec:numimp}).
Increasing the number of momentum bins from the canonical 100 to 150 bins (thick black dashed and dotted lines) improves the situation, although it is clear that we still do not have perfect detailed balance.
Nonetheless, the induced error is small enough that we can conclude it is under control.


\subsection{Comparison of approximation schemes}
\label{sec:comparison}

The different approximation schemes have different strengths and weaknesses as already discussed in \sref{sec:approx_intro}. On the one hand, the equilibrium approximation has the advantage that the assumptions involved have been analysed quite extensively~\cite{Bell:1998ds}. It also allows the scattering processes to push the distributions towards the equilibrium form as they should, albeit at the cost of sacrificing detailed balance.
The CC approximation on the other hand enforces detailed balance as it prevents the same scattering processes from repopulating the active neutrino states.  However, in doing so, the approximation also neglects  possible refilling due to redistribution between different momentum states. 
With these considerations in mind, we expect the equilibrium approximation to overestimate the neutrino number and energy densities, and the CC approximation to underestimate them.  

These trends are reinforced and possibly enhanced by the behaviour of the damping term in the two approximation schemes. By neglecting Pauli blocking, the equilibrium approximation overestimates the damping, while the CC approximation underestimates it through a missing  positive Pauli blocking term discussed in \sref{sssec:damping}. 
The size of the damping term has direct consequences for $\dNeff$ as the production of the sterile neutrinos in our case is non-resonant, where the effective production rate,
\begin{align}
\label{eq:effectiveprodrate}
\Gamma = \frac{1}{2}\sin^2(2\theta_m) \Gamma_{\textrm{collision}},
\end{align}
is directly proportional to the damping term  $\Gamma_{\textrm{collision}} \sim D$ and the in-matter mixing angle~$\theta_m$~\cite{Kainulainen:1990ds}. 
Thus, the larger the damping, the higher the resulting $\Delta N_{\rm eff}$, and vice versa.

\begin{figure}%
\center
\includegraphics[width=0.79\columnwidth]{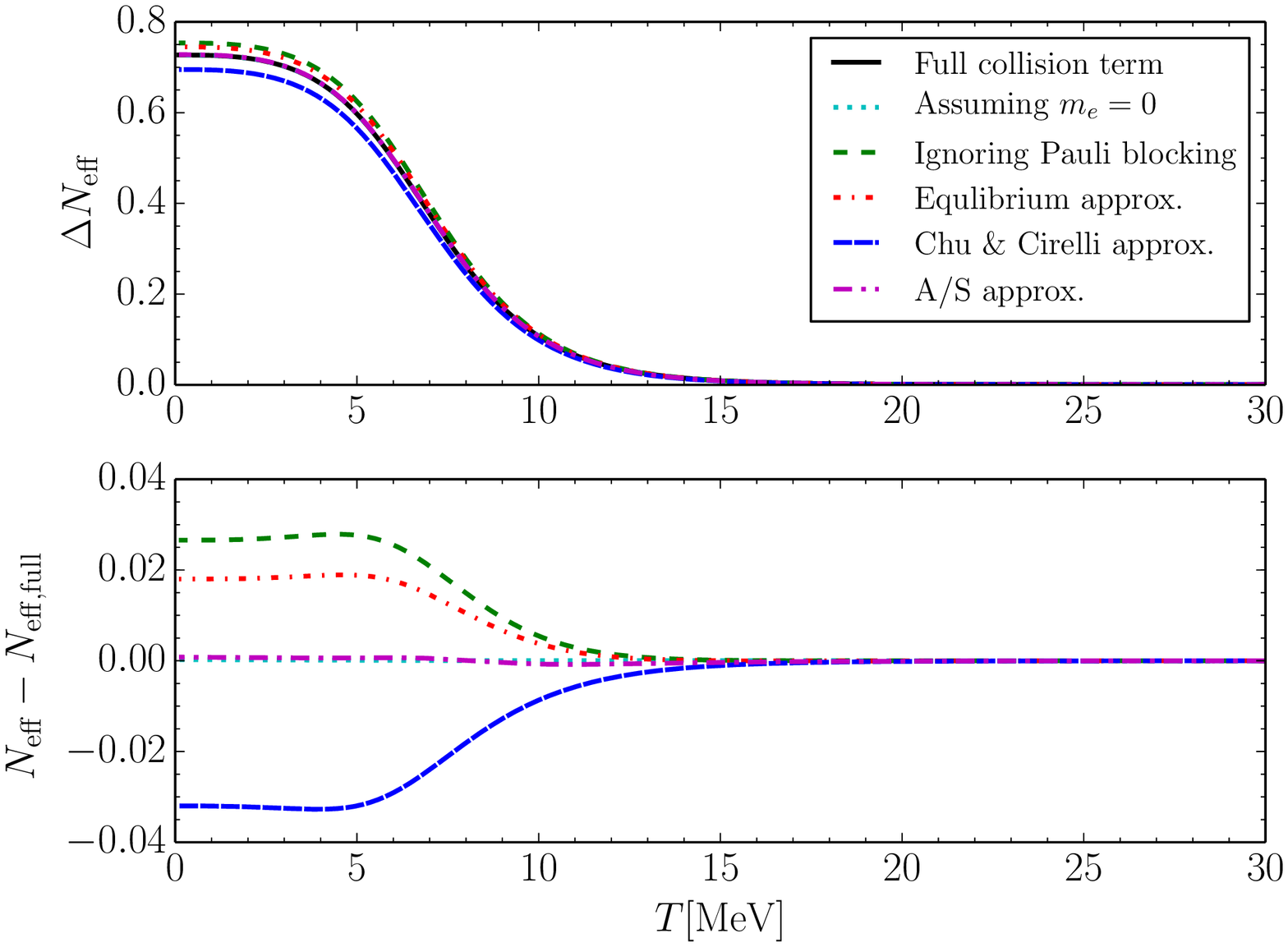}%

\includegraphics[width=0.79\columnwidth]{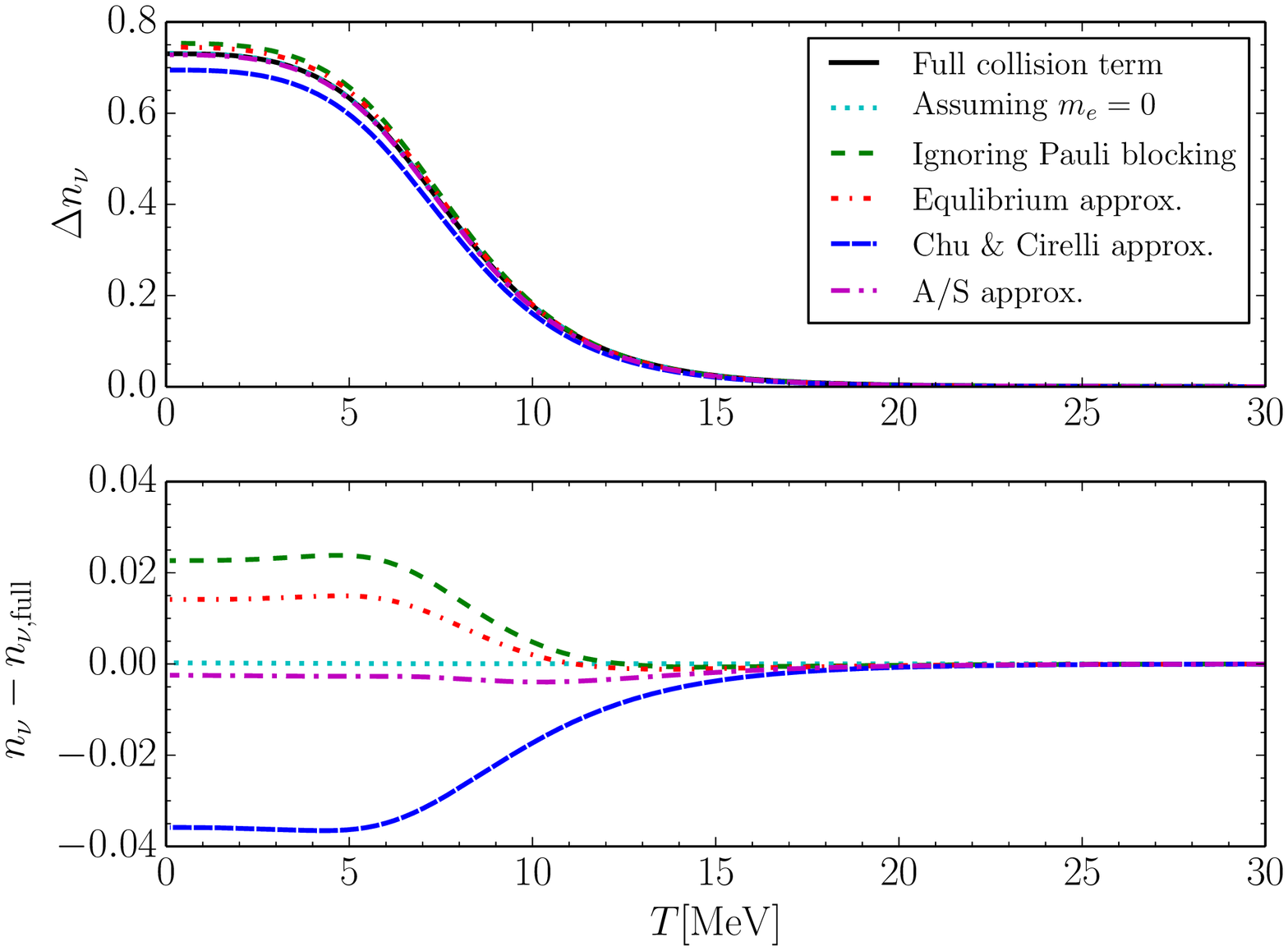}%
\caption{Comparison of the full treatment (black solid lines) with the equilibrium approximation (red dot-short dash), the CC approximation (blue long dashed), and the A/S approximation (purple dot-long dash) introduced in this work.
Setting $m_e=0$ (cyan dotted) has almost no effect, while ignoring Pauli blocking in the full expression (green dashed) gives almost the same result as the equilibrium approximation.}%
\label{fig:Nn}%
\end{figure}

As shown in \fref{fig:Nn}, the over- and underestimation of $\Delta n_\nu$ and $\Delta N_{\rm eff}$ in the equilibrium and CC approximations, respectively, are exactly what transpires in our numerical solutions of the QKEs  for the benchmark mixing parameter values.   In contrast, 
the A/S approximation introduced in this work takes the best of both worlds, and, as is evident in \fref{fig:Nn}, comes much closer to the full collision treatment than both the equilibrium and the CC approximations.   The A/S approximation is also expected to overestimate somewhat the damping relative to the full treatment, since 
oscillations generally push the distribution functions below the equilibrium values.  This is however a very small effect, and the fact that the resulting $\Delta N_{\rm eff}$ and 
$\Delta n_\nu$ values tend towards different sides of the full solutions suggests that the A/S damping term is not the cause of any major systematic bias.

\subsection{Electron mass and Pauli blocking}

Next we test the consequences of ignoring Pauli blocking, and of assuming $m_e=0$ in the full collision term. As shown in \fref{fig:Nn}, the latter assumption makes no practical difference to $\Delta n_\nu$ and $\Delta N_{\rm eff}$, since the conversion to sterile neutrinos takes place largely before electrons become nonrelativistic. 
Ignoring  Pauli blocking, however, induces a $\sim 0.02$  absolute error as $T$ drops below $\sim 5$~MeV, smaller than the naive expectation of $\sim 10\%$ estimated from the Pauli blocking terms in the relevant momentum range~\cite{Bell:1998ds}.
 
 Note that there is a small subtlety when ignoring Pauli blocking: detailed balance can be satisfied for Fermi--Dirac distributions only when the Pauli blocking is taken into account.  Otherwise, ignoring Pauli blocking generally demands that we replace Fermi--Dirac with Maxwell--Boltzmann statistics in order to fulfil detailed balance. We would however like to continue using Fermi--Dirac distributions, and therefore choose to enforce detailed balance  in a different way. For all processes of the form $a(p) + \nu_\alpha(k) \rightleftarrows b(p') c(k')$, we assume that
\begin{equation}
  \label{eq:noP}
  f(E_{p'}) f(E_{k'}) - f(E_p)f(k) = \feq(E_p) (\feq(k) - f(k)).
\end{equation}
This is also the assumption behind the equilibrium approximation, and is exact if $f(E_{p'})$, $f(E_{k'})$ and $f(E_p)$ take on the Maxwell--Boltzmann equilibrium form.
Not surprisingly we see in \fref{fig:Nn} that the equilibrium approximation solution follows the no Pauli blocking full solution quite well; the difference between them is
due mainly to rounding errors in the numerical value of $C_\alpha$ in the equilibrium approximation immediately below equation~\eref{eq:approxD}.


\subsection{Impact on distribution functions}

\begin{figure}%
\center
\includegraphics[width=\columnwidth]{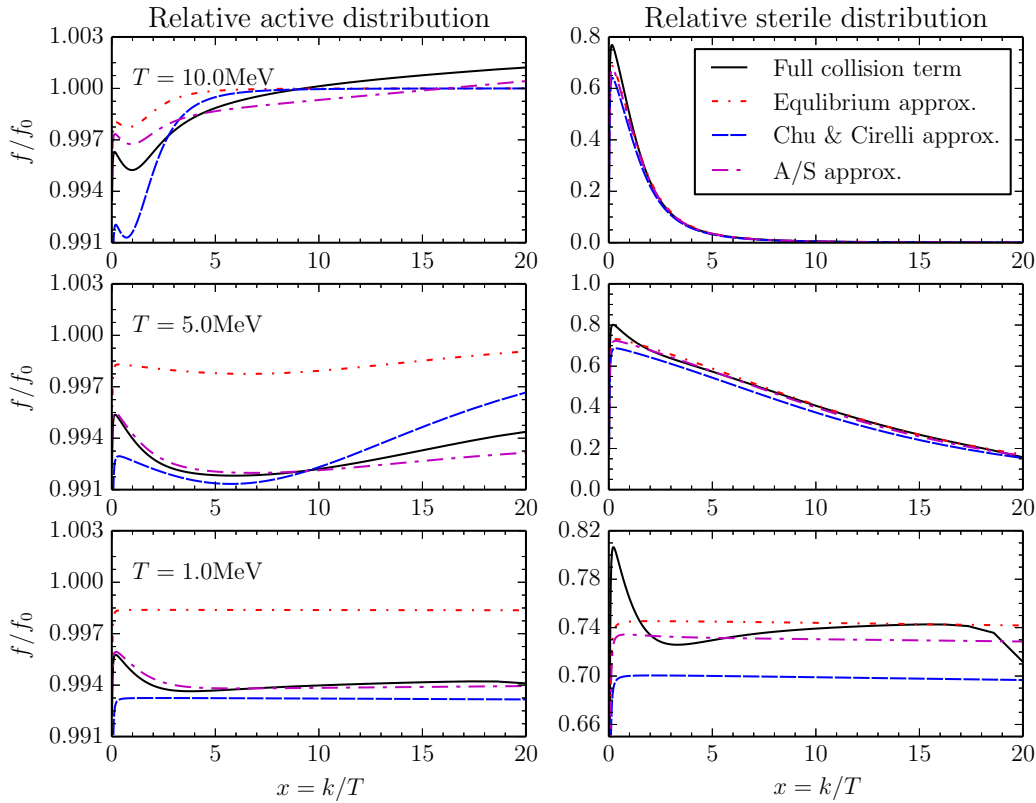}%
\caption{The active and sterile neutrino momentum distributions computed from the full treatment (black solid lines), the equilibrium approximation (red fit-short dash), the CC approximation (blue dashed), and the A/S approximation (purple dot-long dash) for the benchmark mixing parameters $\delta m^2 = 0.1\electronvolt^2$ and $\sin^2(2\theta) = 0.025$ at three different temperatures.  The distributions have been normalised to the relativistic Fermi--Dirac distribution~$f_0$.}%
\label{fig:spectra}%
\end{figure}

It is also interesting to compare the active and sterile neutrino distribution functions in the different approximations. The distribution function of the electron neutrino affects directly  the weak reaction rates in BBN~\cite{Saviano:2013ktj}, while the division of $\dNeff$ between $\nu_\alpha$ and $\nu_s$ have consequences for large-scale structure formation if $\delta m^2$ is significantly larger than the solar and atmospheric mass splittings~\cite{Mirizzi:2014ama, Chu:2015ipa}.
Figure~\ref{fig:spectra} shows this comparison for the benchmark mixing parameters.

The first observation is that  the full collision treatment  gives rise to $f>f_0$ for the high-momentum active neutrinos at $T=10~\mega\electronvolt$.  This is not a numerical artefact, but originates in both the $\nu_\alpha\nu_\alpha$-scattering and the annihilation processes in the presence of a step-like feature at low momenta, such as that produced by oscillations here.  Consider first $\nu_\alpha\nu_\alpha$-scattering.  Removing neutrinos from the low momentum modes causes the energy per neutrino to increase.  This in turn triggers the number and energy conserving scatterings to push the distribution towards an equilibrium with a higher effective temperature, which automatically leads to $f>f_0$. 
   Annihilation processes on the other hand enhance $f$ via a slightly different mechanism.  For processes involving only states above the step, the equilibrium is maintained.  For processes such as $\nu_\alpha(k_{\textrm{high}}) \bar\nu_\alpha(p_{\textrm{low}}) \rightleftarrows a(k') \bar a(p')$, however, the reaction will be pushed to the left because there is a deficit of $\bar\nu_\alpha(p_{\textrm{low}})$ relative to $a(k')$ and $\bar a(p')$. Thus, annihilation can also overproduce $\nu_\alpha$ at high momenta, leading to $f > f_0$.

\begin{figure}%
\center
\includegraphics[width=0.5\columnwidth]{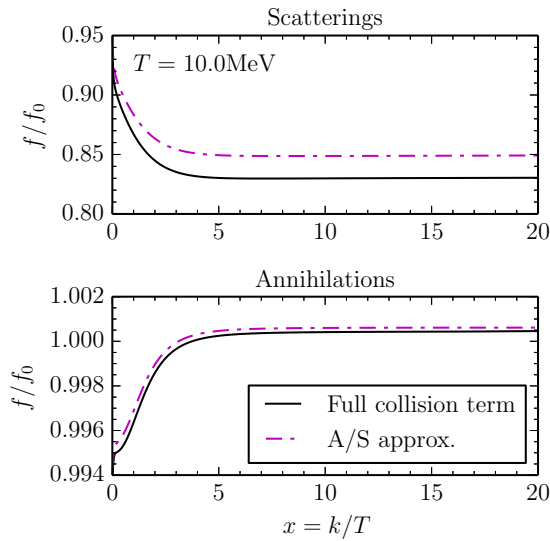}%
\caption{The active momentum distributions at $T=10$~MeV  calculated using a collision term incorporating only scattering processes ({\it top}) and only annihilations ({\it bottom}). The difference between the full treatment (black solid) and the A/S approximation (purple long-dot dash) is much larger for the scatterings than for the annihilations. We have used the benchmark mixing parameters $\delta m^2=0.1\electronvolt^2$ and $\sin^2(2\theta)=0.025$, and the distributions have been normalised to the relativistic Fermi--Dirac distribution $f_0$.}%
\label{fig:scatanni}%
\end{figure}

The A/S approximation mimics both the annihilation and the scattering effects to an extent, and is the only approximation scheme that manages to reproduce $f> f_0$ at $T=10$~MeV, albeit  only at very high momenta.
The annihilation effect is captured by the $n_{\nu_\alpha}$~factor in equation~(\ref{eq:Ras}), while the separate treatment of $\nu_\alpha \pbar{\nu}_\alpha \rightleftarrows \nu_\alpha \pbar{\nu}_\alpha$ accounts for the scattering effect. The latter constitutes the main role of the $C_{\alpha,\nu}$ term in the A/S repopulation approximation~(\ref{eq:Ras}); in contrast, the $C_{\alpha, s}$ term with the pseudo-chemical potential~$\xi$ as the sole degree of freedom acts in the wrong direction: it becomes negative when the active sector is depleted, giving rise to a lower value of $f$  for any choice of momentum.

Notwithstanding its success at capturing some degree of the $f>f_0$ phenomenon, the approximate treatment of scattering remains  the main source of discrepancy between the full solution and the A/S approximation at  $T = 10~\mega\electronvolt$. This is evident in figure~\ref{fig:scatanni} where we compare the distribution functions obtained assuming (i) only scattering, and (ii) only annihilation.  The scattering-only result also demonstrates that these processes do not replenish the active neutrino population, yielding $f/f_0 \sim 0.85$ for most momenta, while the annihilations are very effective at bringing $f/f_0$ to 1.

Apart from these effects, we find that the equilibrium approximation gives too large a final distribution for both the active and the sterile neutrinos,  as expected from the oversized $R_{\rm eq}$ and $D_{\rm eq}$ terms.
 The CC approximation, on the other hand, comes surprisingly close to the real active neutrino distribution, but is short by $\sim 5$\% for the sterile states. As repopulation does not directly affect oscillations into sterile states, we conclude that this offset must originate in the undersized damping term~$D_{\rm CC}$, which as discussed in section~\ref{sec:comparison} affects directly the effective sterile neutrino production rate~\eref{eq:effectiveprodrate}.


\subsection{Dependence on mixing parameters}

So far we have tested the various approximation schemes using the benchmark mixing parameter values $\delta m^2 = 0.1~{\rm eV}^2$ and  $\sin^2 2 \theta = 0.025$.
In reality, however, the errors caused by these approximations are also sensitive to the parameter values.

\begin{figure}%
\center
\includegraphics[width=0.6\columnwidth]{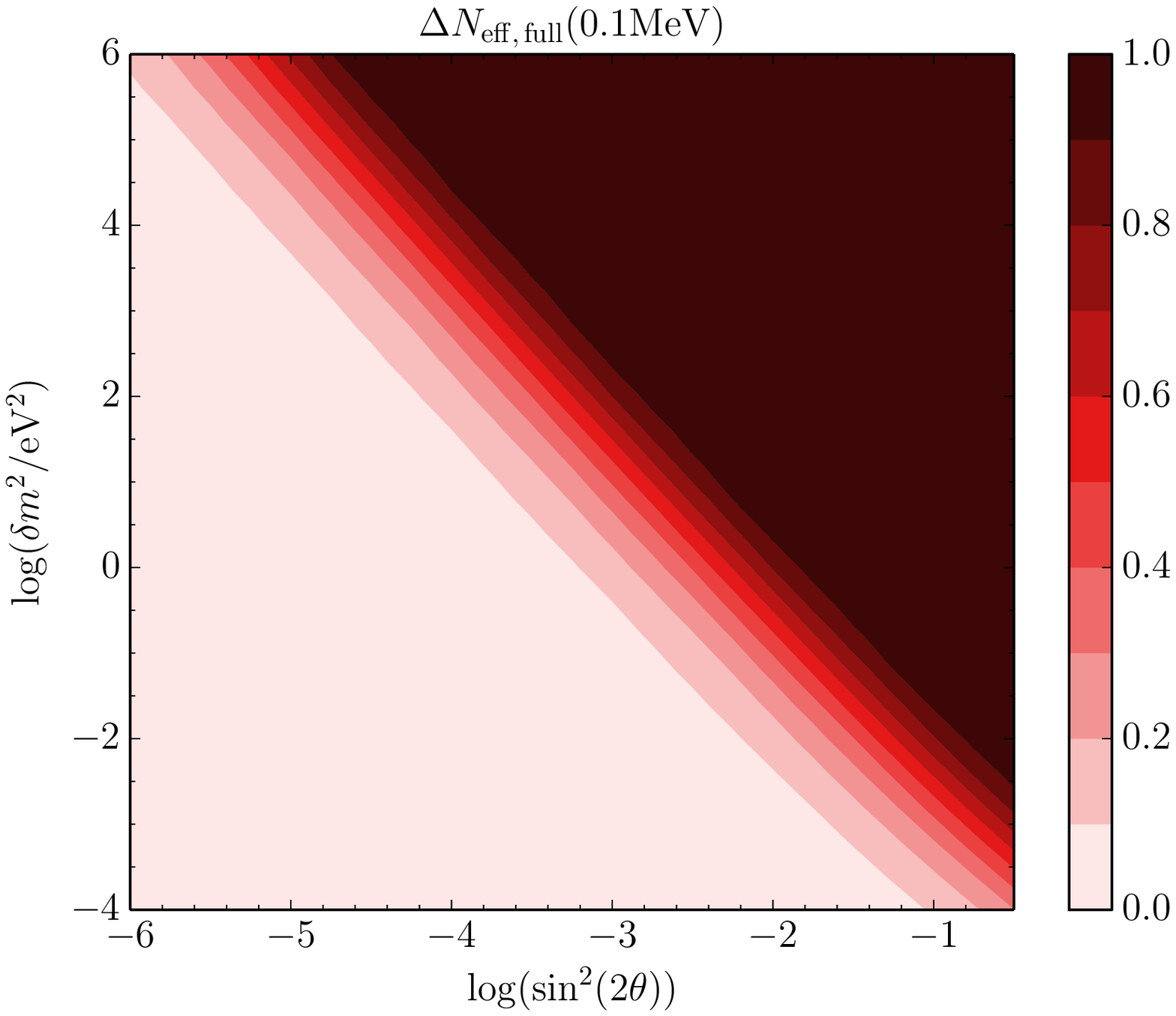}%
\caption{$\dNeff$ at $T = 0.1~\mega\electronvolt$ using the full collision term.}%
\label{fig:full_grid}%
\end{figure}

\begin{figure}[tbp]%
\center
\includegraphics[width=0.48\columnwidth]{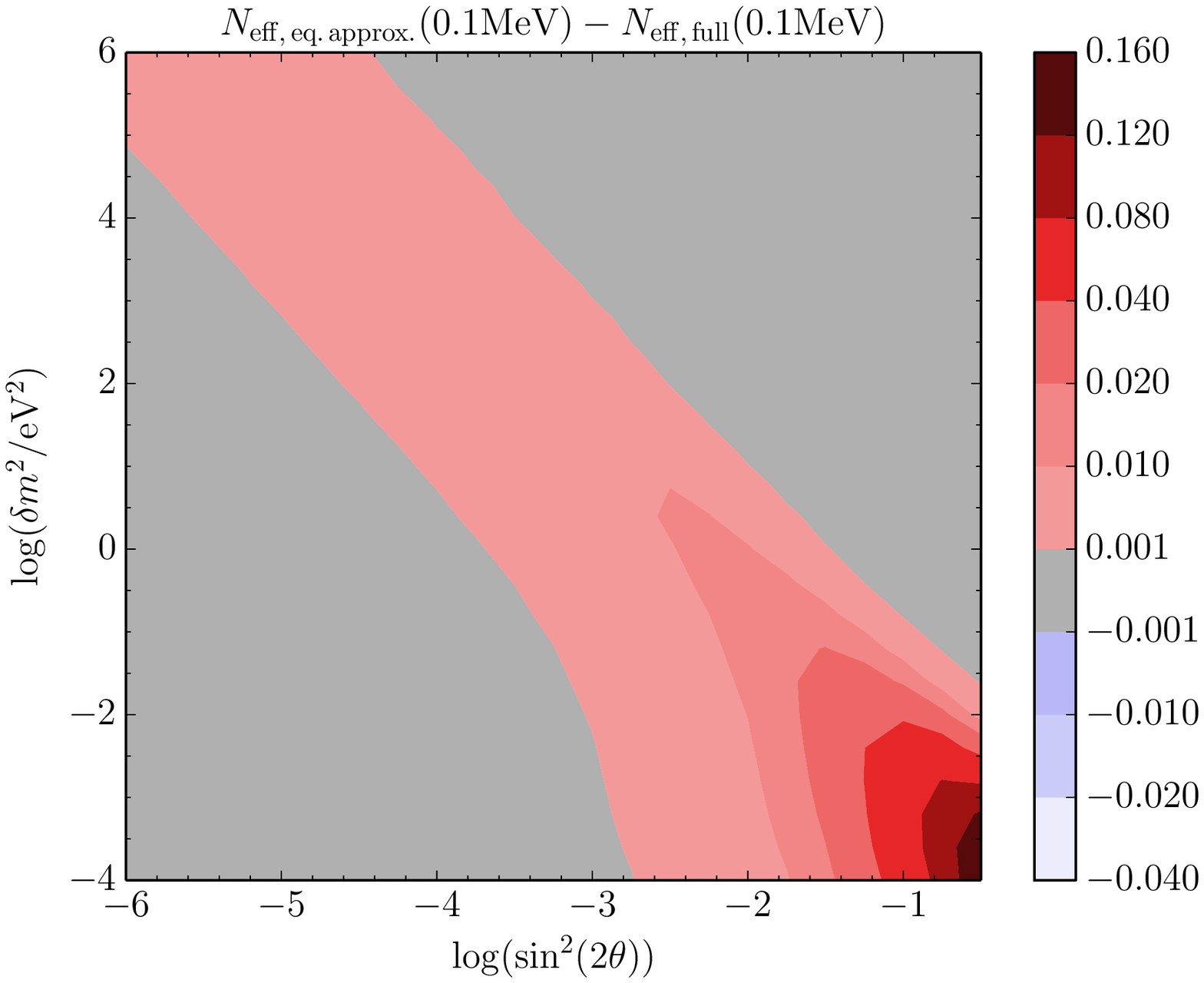}%
\includegraphics[width=0.48\columnwidth]{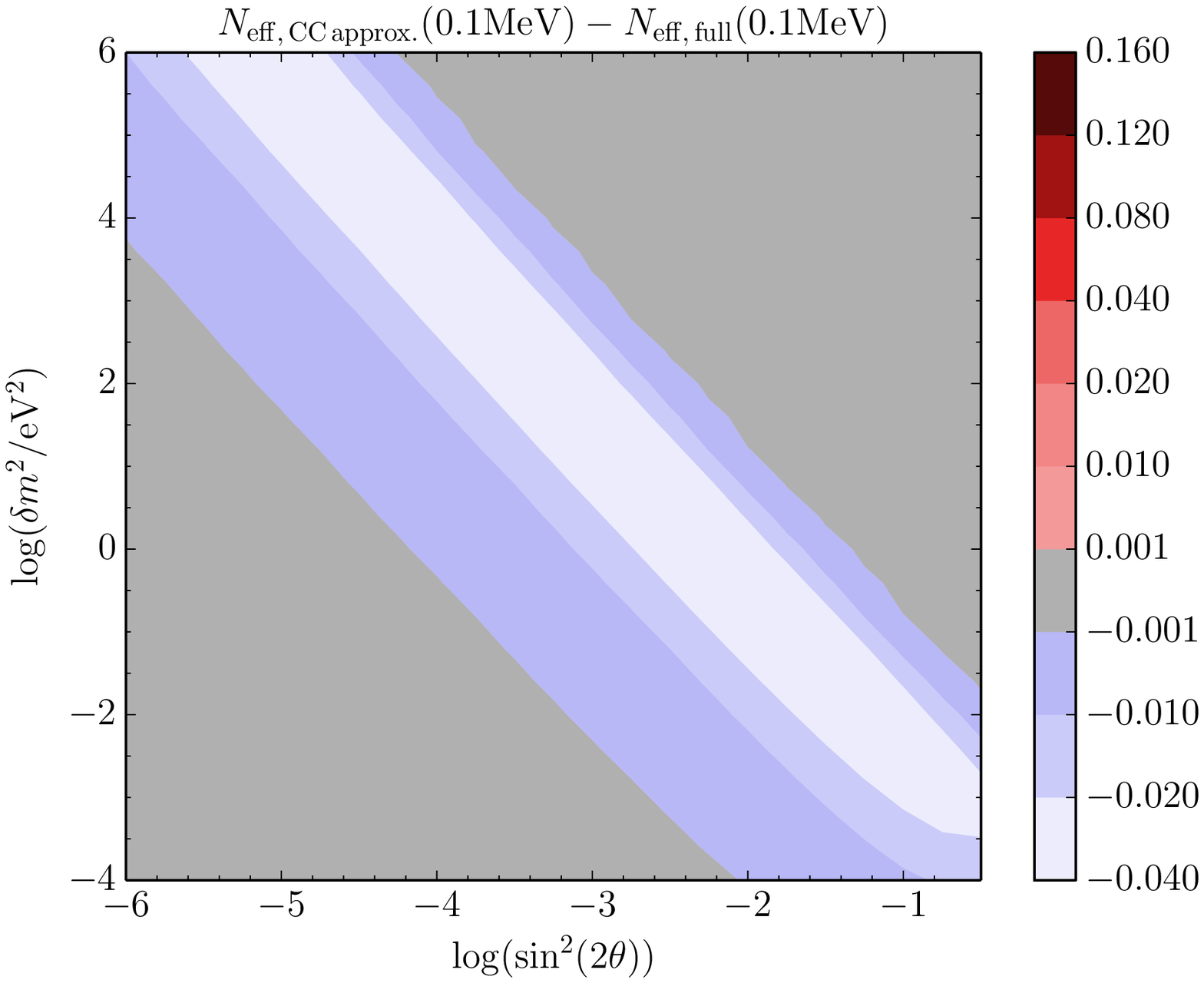}\\%
\includegraphics[width=0.48\columnwidth]{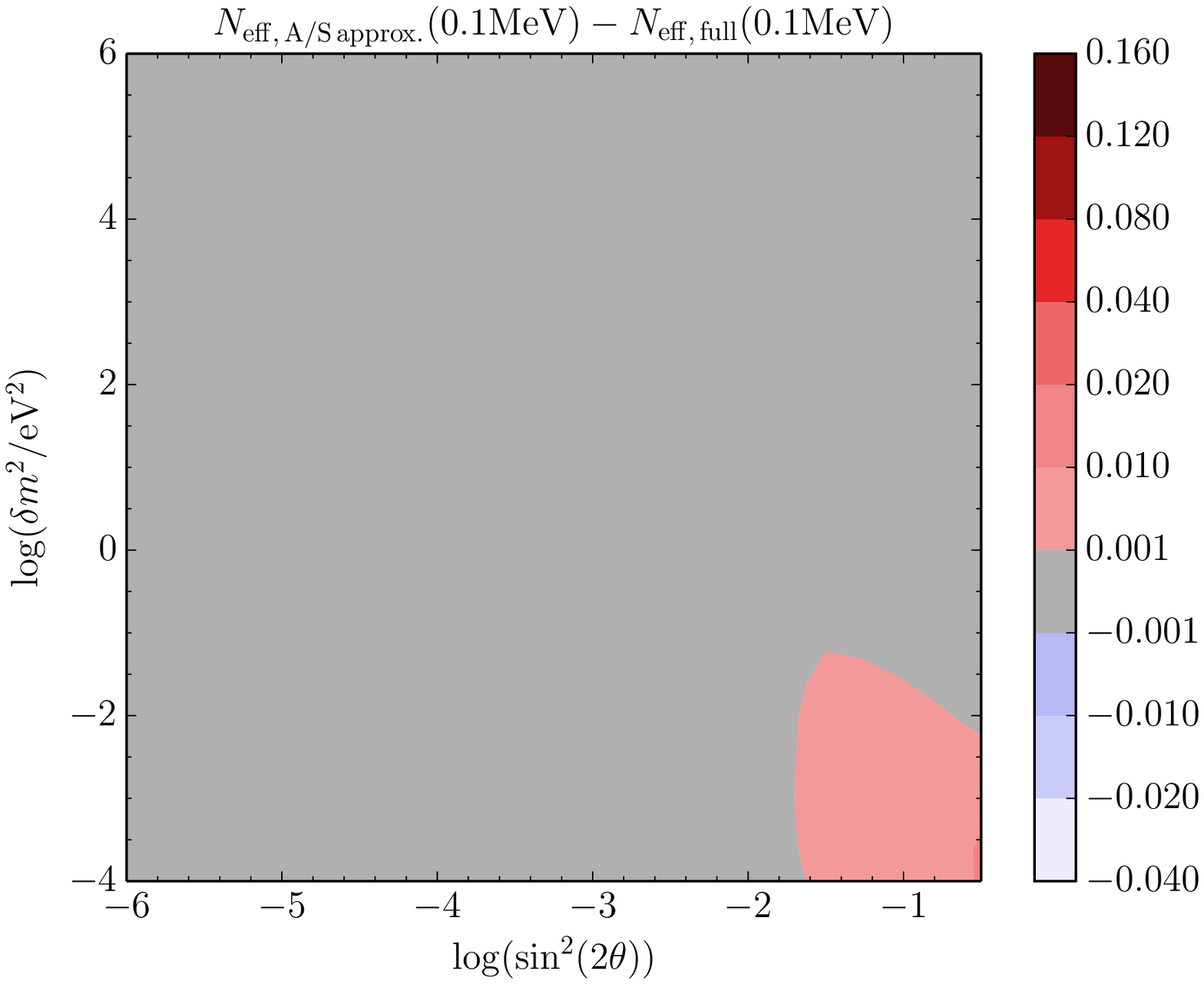}%
\caption{Deviations in the $\dNeff(T = 0.1~{\rm MeV})$ values  computed using various approximations from the full solutions.  {\it Top left:} The equilibrium approximation.  {\it Top right:} The CC approximation.  {\it Bottom:} The A/S approximation.
Note that the colour scale is linear, but the contour levels are not.}%
\label{fig:approx_grid}%
\end{figure}

Figure~\ref{fig:full_grid} shows $\Delta N_{\rm eff}$ at $T =0.1$~MeV as a function of $\delta m^2$ and $\sin^2 2 \theta$ computed using the full collision term. 
Our results are generally quite similar to previous calculations using the equilibrium approximation~\cite{Hannestad:2012ky, Enqvist:1991qj}.  Note however that for large mixing angles and small mass squared differences, the contours deviate slightly from the straight lines found in~\cite{Hannestad:2012ky, Enqvist:1991qj}. To highlight this deviation, we show in  \fref{fig:approx_grid} the differences in $\Delta N_{\rm eff}$ incurred respectively by the equilibrium, CC, and A/S approximations relative to the full solution in the $(\delta m^2, \sin^2 2 \theta)$-parameter space.

As expected, the deviations always occur, independently of the approximation scheme, in a diagonal band corresponding to the transition region from $\dNeff = 0$ to $\dNeff=1$.  
Beyond this common feature, however, 
the different approximations incur the largest deviations at different parameter values.

 The equilibrium approximation follows the result of the full calculation quite well for $\delta m^2$ values above $0.01 \electronvolt^2$, but overestimates $\dNeff$ by more than 0.1 at $\delta m^2 < 0.001~\electronvolt^2$. We can understand this deviation by looking at the conversion temperature. The temperature of maximal conversion is proportional to $(\delta m^2)^{1/6}$~\cite{Enqvist:1991qj}, so that low values of $\delta m^2$ generally correspond to low conversion temperatures.  If the conversion temperature is sufficiently high ($ T \gg 1$~MeV),
repopulation is rapid and $\dNeff$ is limited only by how fast $\nu_s$ can be produced through oscillations and collisions~\cite{Kainulainen:1990ds}.  
 The effective production rate is given by equation~\eref{eq:effectiveprodrate}, and production ceases as soon as the collision rate becomes too low. 
 If most of the conversion occurs at low temperatures, however, collisions become too inefficient to sustain the population of the active sector and consequently for equation~\eref{eq:effectiveprodrate}---which assumes instantaneous repopulation---to hold, thereby
 causing the real $\Delta N_{\rm eff}$ contours to deviate from straight lines in  relation to $\delta m^2$ and $\sin^2 2\theta$ in  \fref{fig:full_grid}.  
 The equilibrium approximation errs in its overestimation of the repopulation rate, yielding almost straight $\dNeff$  contours even in the low $\delta m^2$, high $\sin^2 2 \theta$ region.  
 
In contrast, the top right panel of \fref{fig:approx_grid} shows that the CC approximation generally underestimates $\dNeff$, but works somewhat better at $\delta m^2 \lesssim 0.01~\electronvolt^2$. For high $\delta m^2$ values the underestimation is due mainly to the undersized $D_{\rm CC}$ damping term which diminishes the sterile neutrino production rate as already discussed in section~\ref{sec:comparison}.  For low $\delta m^2$ values, however, the agreement becomes better (a deviation of $0.02$ at $\delta m^2=10^{-4}$, $\sin^2 2\theta=10^{-0.5}$)  because the deficiency of {\it sterile} neutrinos is compensated by an overproduction of active neutrinos (similar to the overpopulation of the active sector discussed above in relation to the equilibrium approximation).  This overproduction comes about as follows.  Although technically the CC repopulation integral  incorporates only annihilation processes, numerically
the constant $2(\gamma_a^e)^2/3.15 = 0.317$ is rather larger than its annihilation counterpart in the A/S approximation, $C_{e,a} = 0.177$. Thus, 
the CC approximation does inadvertently contain some degree of repopulation due to scattering, which drives up the active neutrino repopulation rate relative to the true rate.

Finally, the bottom panel of \fref{fig:approx_grid} shows the A/S approximation. Here, the agreement is generally much better than either the equilibrium or the CC approximation, although there remains a small discrepancy of $\sim 0.01$ in the  low $\delta m^2$, high $\sin^2 2 \theta$ corner.  This is not surprising, as the region is characterised by large spectral distortions from conversion at low temperatures, and is thus most sensitive to how exactly we handle repopulation in the active sector.
It is nonetheless remarkable that for most of the parameter region, the A/S approximation is able to reproduce the full results at a precision of $\sim 0.001$ through a fairly simplistic description of the collision term. This deviation is in fact comparable to the numerical error expected of the full treatment due to our choice of momentum resolution (see \sref{sec:convergence}).   Therefore, \fref{fig:approx_grid} not only validates the A/S approximation, but through comparison with  a physically intuitive modelling of collisions, also confirms that the full collision term has been implemented correctly


\section{Conclusions}
\label{sec:conclusion}

We have calculated in this work the full collision term for active--sterile neutrino oscillations in the early universe, and for the specific case of $(\nu_e,\nu_s)$-oscillations implemented it in the computation of $\dNeff$ from sterile neutrino thermalisation.  In particular we have included for the first time a nonzero electron mass and Pauli blocking in the collision integrals.
The former turns out to have a negligible impact on the final results, while ignoring the latter gives rise to noticeable discrepancies.  Nonetheless, 
our full treatment confirms previous analyses based on approximations of the collision terms  that a $1~\electronvolt$ sterile neutrino coupled with a mixing angle $\sin^22\theta \sim 0.1$ produces $\dNeff \approx 1$; the discrepancies arising from approximations are at the level that affects only precision calculations.

We then proceeded to  perform a systematic comparison of the full collision treatment with two approximation schemes found in the literature.  We find that the commonly used ``equilibrium approximation''~\cite{Bell:1998ds} reproduces $\dNeff$ at better than 0.04 for $\delta m^2 > 0.01~\electronvolt^2$, but incurs large errors ($>0.1$)
for very small mass squared differences due to the unphysical repopulation of the active sector by elastic scattering. The ``CC approximation'' of Chu and Cirelli~\cite{Chu:2006ua}
is discrepant up to 0.04 for $\delta m^2 > 0.0001~\electronvolt^2$, but improves for low $\delta m^2$ values because of a fortuitous cancellation between an underestimated damping term and an overestimation of  repopulation from annihilation processes.

Recognising that the equilibrium and CC approximations neglect different physical effects, we have devised a new approximation scheme, the A/S approximation, in which scattering and annihilation contributions to repopulation are treated separately, and the damping term includes Pauli blocking.   As the scheme is better able to capture the physics of repopulation and damping, it also pushes the error in $\dNeff$ down to 0.001 for most of the parameter region, although minor deviations  ($\sim 0.01$) remain in the
 $\delta m^2 < 0.001~\electronvolt^2$ region, where the sterile neutrino conversion temperature is lowest and hence  spectral distortions from oscillations are expected to have the largest effect.
 
The connection between low $\delta m^2$ deviations and low temperature spectral distortions also has implications for an inverted mass spectrum, i.e., where the active state is heavier than the sterile state, as well as for the various mechanisms designed to reconcile $\electronvolt$-mass sterile neutrinos with cosmology. In the case of an inverted mass spectrum, sterile neutrino thermalisation proceeds via a resonance, which, depending on the adiabaticity of the resonance and hence the mixing angle, can cause more disturbance to the active neutrino spectrum than in the non-resonant case.  Likewise, mechanisms that block the production of $\electronvolt$-mass sterile neutrinos typically work by delaying  the thermalisation to low temperatures, which by construction also makes them more sensitive to the approximations employed for the collision terms.

At the current level of observational precision---$\sigma(\dNeff) \approx 0.2$ from Planck~\cite{Planck:2015xua}---our analysis shows that the CC and the A/S approximations can be reliably applied to most active--sterile oscillation scenarios, whereas the equilibrium approximation appears to be approaching its boundary of validity if the sterile neutrino conversion temperature is too low.  In the future, large-volume galaxy surveys are expected to improve the sensitivity to $\dNeff$ to $\sim 0.03$~\cite{Basse:2013zua}.
Thus, should hints for a $1~\electronvolt$ sterile neutrino persist in the laboratory, a collision treatment more precise
than either the equilibrium or the CC approximation can offer would become necessary.  Short of evaluating the full collision terms, which is numerically costly or possibly even infeasible in some cases,  the A/S approximation developed in this work appears to be a most convenient alternative.


\section*{Acknowledgments}

We acknowledge use of computational resources provided by the Danish e-Infrastructure Cooperation. Y$^3$W thanks the Institute for Nuclear Theory at the University of Washington for its hospitality and the Department of Energy for partial support during the completion of this work.


\appendix

\section{Derivation of the full collision terms}
\label{sec:full}

\subsection{Neutrino--electron scattering in the $s$-channel}

Consider first the case of $\nu_\alpha + e^-\rightleftarrows \nu_\alpha + e^-$.  The repopulation term due to this process takes on the form
\begin{equation}
\begin{aligned}
  R_{\alpha,s,e}
  =& \frac{1}{2^9 \pi^5}  E_k \int d^3\k'd^3\p'd^3\p\:  \delta^{(4)}(k+p-k'-p') \frac{1}{E_{k'}E_pE_{p'}}\\
 & \times |M|^2(\nu_\alpha(k),e^- (p)| \nu_\alpha(k'), e^-(p')) \left[f_{\nu_\alpha}(E_{k'})f_e(E_{p'})(1-f_{\nu_\alpha}(E_k)) (1-f_e(E_p))\right. \\
 &\left. \hspace{6.cm} - f_{\nu_\alpha}(E_k) f_e(E_p)(1-f_{\nu_\alpha}(E_{k'}))(1-f_e(E_{p'}))\right],
 \label{eq:rse}
\end{aligned}
\end{equation}
where the matrix element is~\cite{Flaig:1989,Hannestad:1995rs} 
\begin{equation}
  |M|^2 = 32 G_F^2 \left(A_\alpha^2 (k\cdot p)(k'\cdot p') + B_\alpha^2 (k\cdot p') (k'\cdot p) - m_e^2 A_\alpha B_\alpha (k \cdot k')\right),
  \label{eq:matrix}
\end{equation}
with $A_e = 2\sin^2\theta_W + 1$, $A_{\mu, \tau} = 2\sin^2\theta_W - 1$, $B_e = 2\sin^2\theta_W$, and $B_{\mu,\tau} = 2\sin^2\theta_W$.
Note the change of notation here in the appendix: $p$  and $\p$ now denote respectively the 4- and 3-momentum, while  in the main text we use $p$ to indicate the magnitude of the 3-momentum.

The matrix element has three different dependences on the momentum, and therefore requires three different parameterisations to simplify the integral. We label 
the first dependence $(k\cdot p)(k' \cdot p')$ the $s$-channel,  the second $(k\cdot p')(k' \cdot p)$ the $u$-channel, and the last dependence $(k\cdot k')$ the $t$-channel,
where for each channel it is convenient to define a 3-momentum variable, given respectively by
 \begin{align*}
 \q &\equiv \p+\k = \p'+\k', \\
 \v &\equiv \p - \k' = \p'-\k ,\\ 
 \w & \equiv \k - \k' = \p'-\p,
 \end{align*}
 which replaces one of the integration variables of equation~(\ref{eq:rse}).  Then, to evaulate the integral for each channel we simply follow the technique described by Hahn-Woernle, Pl\"umacher and Wong in~\cite{HahnWoernle:2009qn}.

Taking the $s$-channel as a worked example, we use the 3-momentum variable $\q$ as a reference and define around it a coordinate system
\begin{align*}
 \q &= |\q| (0,0,1),\\
 \k &= E_{k} (0,\sin\eta,\cos\eta),\\
 \k'&= E_{k'} (\cos\phi \sin\theta, \sin\phi \sin\theta, \cos\theta).
\end{align*}
With this choice, we find the quantities
\begin{align*}
 s &= (p+k)^2 = (p'+k')^2 = (E_p+E_k)^2 - |\q|^2,\\
 k\cdot p &= k'\cdot p'= \frac{s-m_e^2}{2},\\
 |\q-\k'|^2 &= |\q|^2 + |\k'|^2 - 2 \q\cdot\k' = |\q|^2 + E_{k'}^2 - 2 |\q| E_{k'} \cos\theta,\\
 |\q-\k|^2 &= |\q|^2 + |\k|^2 - 2 \q\cdot\k = |\q|^2 + E_{k}^2 - 2 |\q| E_{k} \cos\eta,
\end{align*}
and consequently
\begin{equation}
  |M_s|^2 = 8 \: G_F^2 A_\alpha^2 ((E_p+E_k)^2-|\q|^2 - m_e^2)^2
  \label{eq:matrixs}
\end{equation}
for the first term of the matrix element~\eref{eq:matrix}.

Since the matrix element~\eref{eq:matrixs} now  depends only on energies and the magnitude of $\q$, we can use the Dirac delta functions in the integral~\eref{eq:rse} to integrate out the directional dependencies.  Evaluating first the $\p'$-integral as~\cite{HahnWoernle:2009qn}
\begin{equation}
\begin{aligned}
 \int \frac{d^3\p'}{2E_{p'}} \delta^{(4)}(k+p-k'-p') 
=& \int d^3\p'dE_{p'} \frac{\delta(E_{p'}-\sqrt{|\p'|^2+m_e^2})}{2\sqrt{|\p'|^2+m_e^2}} \theta(E_{p'}-m_e)\\
& \times \delta(E_k+E_p-E_{k'}-E_{p'}) \delta^3(\k + \p - \k' - \p')\\
=& \frac{\delta(E_k+E_p-E_{k'} - \sqrt{|\q-\k'|^2+m_e^2})}{2 \sqrt{|\q-\k'|^2+m_e^2}} \theta(E_k+E_p-E_{k'}-m_e),
\label{eq:diracdelta}
\end{aligned}
\end{equation}
we use the property
\begin{align*}
  \delta(f(x)) = \sum_{f(x_i) = 0} \frac{\delta(x - x_i)}{|f'(x_i)|}
\end{align*}
and hence
\begin{equation*}
  \delta(E_i^2-|\p_i|^2 - m_i^2) = \frac{\delta(E_i-\sqrt{|\p_i|^2+m_i^2})}{2\sqrt{|\p_i|^2+m_i^2}} + \frac{\delta(E_i+\sqrt{|\p_i|^2+m_i^2})}{2\sqrt{|\p_i|^2+m_i^2}}
\end{equation*}
to further simplify equation~(\ref{eq:diracdelta}) to~\cite{HahnWoernle:2009qn}
\begin{equation}
\begin{aligned}
 & \int \frac{d^3\p'}{2E_{p'}} \delta^{(4)}(k+p-k'-p') \\
 & \hspace{10mm}= \delta((E_k+E_p-E_{k'})^2 - |\q-\k'|^2-m_e^2)\, \theta(E_k+E_p-E_{k'}-m_e)\\
& \hspace{10mm} = \delta\left((E_k+E_p-E_{k'})^2 - |\q|^2 - E_{k'}^2 + 2 |\q| E_{k'} \cos\theta - m_e^2\right) \, \theta(E_k+E_p-E_{k'}-m_e)\\
& \hspace{10mm} = \frac{1}{2|\q|E_{k'}} \delta \left(\cos\theta - \frac{E_{k'}^2 - (E_k+E_p-E_{k'})^2 + |\q|^2 + m_e^2}{2|\q|E_{k'}}\right)  \, \theta(E_k+E_p-E_{k'}-m_e).
\label{eq:delta1}
\end{aligned}
\end{equation}
In a similar way, we rewrite
\begin{equation}
\begin{aligned}
  \frac{d^3\p}{2E_p} &= \int d^3\p \: dE_p \: \delta(E_p^2-|\q-\k|^2-m_e^2)\: \theta(E_p -m_e)\\
\label{eq:delta2}
&=  \int d^3\q \:  dE_p\: \frac{1}{2|\q|E_k} \: \delta \left( \cos\eta - \frac{E_k^2 - E_p^2+|\q|^2 + m_e^2}{2|\q|E_k}\right)\: \theta(E_p -m_e),
\end{aligned}
\end{equation}
where in the second line we have also changed the integration variable from $\p$ to $\q$.
Then, applying equation~\eref{eq:delta1} and~\eref{eq:delta2} to the integral~\eref{eq:rse}, we find, after performing the trivial angular integrations and averaging over the direction of the incoming neutrinos ($\int d\cos\eta/2$), the $s$-channel contribution
\begin{equation}
\begin{aligned}
  R_{\alpha,s,e,s} =& \frac{1}{2^7 \pi^3E_k^2}\int d\cos\eta \: d\cos\theta \: d|\q| \: dE_{k'} \: dE_p \: \delta(\cos\theta - \dots) \delta(\cos\eta - \dots) \\
& \times  |M_s|^2 F \theta(E_k+E_p-E_{k'}-m_e) \theta(E_p-m_e),
\label{eq:rses}
\end{aligned}
\end{equation}
where $F$ denotes all the distribution functions, and the arguments of the two Dirac delta functions can be read off equations~\eref{eq:delta1} and \eref{eq:delta2} respectively.

The integrals over $\cos\theta$ and $\cos\eta$ involve only Dirac delta functions. Taking the $\cos\eta$-integral as an example and noting that integration limits can be equivalently expressed as step functions, we find
\begin{align*}
 & \int_{-1}^1 d(\cos\eta) \delta\left(\cos\eta - \frac{E_k^2 - E_p^2+|\q|^2 + m_e^2}{2|\q|E_k}\right) \\
 & \hspace{20mm}= \theta\left(1 - \frac{E_k^2 - E_p^2+|\q|^2 + m_e^2}{2|\q|E_k}\right)
\: \theta\left(\frac{E_k^2 - E_p^2+|\q|^2 + m_e^2}{2|\q|E_k}+1\right).
\end{align*}
The two step functions can be reinterpreted as limits on $|\q|$, and together they confine $|\q|$ to the interval
\begin{equation*}
||\k| - |\p|| =  |E_k-\sqrt{E_p^2-m_e^2}| \le |\q| \le E_k + \sqrt{E_p^2-m_e^2} = |\k| + |\p|.
\end{equation*}
Integrating over $d\cos\theta$ gives the same formal result save for the replacements $\k \to \k'$ and $\p \to \p'$.  Thus the combined integration limits on $|\q|$ can be written equivalently as
\begin{equation}
\max\left(||\k|-|\p||,||\k'|-|\p'||\right) \le |\q| \le \min\left(|\k|+|\p|,|\k'|+|\p'|\right),
\label{eq:qlimits}
\end{equation}
and we note that $|\p'|$ is determined from $|\p|,|\k|,|\k'|$ by imposing energy conservation.
Applying the limits~\eref{eq:qlimits} to the integral~\eref{eq:rses} then yields
\begin{equation}
\label{eq:Ralphases}
  \begin{aligned}
  R_{\alpha,s,e,s} =& \frac{1}{2^7 \pi^3E_k^2}\int_{0}^\infty dE_{k'} \int_{\max(m_e,k'-k+m_{e})}^\infty dE_p \int d|\q|  \:  |M_s|^2 F\\
  &\times \theta(|\q| - \max(||\k|-|\p||,||\k'|-|\p'||)) \theta(\min(|\k|+|\p|,|\k'|+|\p'|)-|\q|)
  \end{aligned}
\end{equation}
as our reduced repopulation integral from the $s$-channel.  The $u$- and $t$-channel integral reduction proceeds in a similar manner, using $\v$ and $\w$ respectively as an integration variable.


\subsection{The massive case}
\label{sec:massive_int}

The reduced integral~\eref{eq:Ralphases}  is but one of three contributions to the repopulation of $\nu_\alpha$ arising from neutrino scattering with electrons. The full collision term, including scattering and annihilation processes with $e^{\pm},\pbar{\nu}_\alpha, \pbar{\nu}_\beta$, has in total 14 such terms to be evaluated (see~\tref{tab:matrixel}).
Fortunately, however,  these 14 terms can all be recast into one of the standard $s$-, $t$-, and $u$-forms, and thus can be handled in ways similar to that discussed above.

\begin{table}[t]
  \centering
{\small
  \begin{tabular}{l @{\qquad} l}
    Reaction ($\alpha \neq \beta$)   & \multicolumn{1}{c}{$S|M|^2$}\\\hline
    $\nu_\alpha(k) \nu_\beta(p) \rightarrow \nu_\alpha(k') \nu_\beta(p')$ & $32 G_F^2 (k\cdot p)(k'\cdot p')$ \\[5pt]
    $\nu_\alpha(k) \bar\nu_\beta(p) \rightarrow \nu_\alpha(k') \bar\nu_\beta(p')$ & $32 G_F^2 (k\cdot p')(k'\cdot p)$ \\[5pt]
    $\nu_\alpha(k) \nu_\alpha(p) \rightarrow \nu_\alpha(k') \nu_\alpha(p')$ & $32 G_F^2 2 (k\cdot p)(k'\cdot p')$ \\[5pt]
    $\nu_\alpha(k) \bar\nu_\alpha(p) \rightarrow \nu_\alpha(k') \bar\nu_\alpha(p')$ & $32 G_F^2 4 (k\cdot p')(k'\cdot p)$ \\[5pt]
    $\nu_\alpha(k) e^-(p) \rightarrow \nu_\alpha(k') e^-(p')$ & $\begin{matrix} 32 G_F^2 \left((2x_W \pm 1)^2(k\cdot p)(k'\cdot p') + 4x_W^2(k\cdot p')(k'\cdot p)\right.\\ \left. \hspace{3cm} - (2x_W \pm 1)2x_Wm_e^2(k\cdot k')\right)\end{matrix}$\\[5pt]
    $\nu_\alpha(k) e^+(p) \rightarrow \nu_\alpha(k') e^+(p')$ & $\begin{matrix} 32 G_F^2 \left((2x_W \pm 1)^2(k\cdot p')(k'\cdot p) + 4x_W^2(k\cdot p)(k'\cdot p') \right.\\ \left. \hspace{3cm} - (2x_W \pm 1)2x_Wm_e^2(k\cdot k')\right)\end{matrix}$\\[5pt]
    $\nu_\alpha(k) \bar\nu_\alpha(p) \rightarrow \nu_\beta(k') \bar\nu_\beta(p')$ & $32 G_F^2 (k \cdot p')(k'\cdot p)$\\[5pt]
    $\nu_\alpha(k) \bar\nu_\alpha(p) \rightarrow e^-(k') e^+(p')$ & $\begin{matrix} 32 G_F^2 \left((2x_W \pm 1)^2(k\cdot k')(p \cdot p') + 4x_W^2(k\cdot p')(k'\cdot p)\right.\\ \left. \hspace{3cm}- (2x_W\pm 1)2x_Wm_e^2(k\cdot p)\right)\end{matrix}$\\[10pt]\hline
  \end{tabular}
  }
  \caption{Matrix elements for all relevant reactions involving $\nu_\alpha$, with $x_W = \sin^2\theta_W = 0.23864$~\cite{Agashe:2014kda}, $S$ is a symmetrisation factor of $1/2$ for each pair of indistinguishable particles in the  initial and the final state, and $|M|^2$ has been summed but not averaged over initial and final spins.  For the process with two $\nu_\alpha$s in the initial state, we have further multiplied the matrix element by $2$ to account for the fact that $\nu_\alpha(k) \nu_\alpha(p) \to \ldots$ and $\nu_\alpha(p) \nu_\alpha(k) \to \ldots$ constitute two identical processes.
Where there is a choice of $\pm$, the plus signs are for $\alpha = e$, and the minus signs for $\alpha = \mu, \tau$. The corresponding matrix elements for $\bar{\nu}_\alpha$ can be obtained by the exchange $(k\cdot p)(k'\cdot p') \leftrightarrow (k\cdot p')(k'\cdot p)$ for the elastic scattering processes.}
  \label{tab:matrixel}
\end{table}

As the reduction procedure concerns only kinematics and does not involve the actual matrix element besides the initial classification of the momentum dependence into $s$-, $u$- or $t$-forms, we shall keep the calculation as general as possible and allow for the possibility that all initial and final states are massive.  Then, the number of independent reductions to be performed is only three, which yield:
\begin{equation}\label{eq:Rs}
\begin{aligned}
  R_{\alpha,\,s\mathrm{-channel}} =&\frac{1}{ 2^7 \pi^3 E_k |\k|}\int_{m_{k'}}^\infty dE_{k'} \int_{\max(m_p,k'-k+m_{p'})}^\infty dE_p \int d|\q| \: S|M_s|^2 F\\
  &\times \theta(|\q|-\max(||\k|-|\p||,||\k'|-|\p'||)) \theta(\min(|\k|+|\p|,|\k'|+|\p'|)-|\q|) ,
\end{aligned}
\end{equation}
\begin{equation}\label{eq:Rt}
\begin{aligned}
  R_{\alpha,\,t\mathrm{-channel}}  
  = & \frac{ A_k}{ 2^7 \pi^3 E_k|\k|} \int_{m_{k'}}^\infty dE_{k'} \int_{\max(m_p,k'-k+m_{p'})}^\infty dE_p \int d|\w| \: S|M_t|^2 F \\
  &\times \theta(|\w|-\max(||\p|-|\p'||,||\k|-|\k'||)) \theta(\min(|\p|+|\p'|,|\k|+|\k'|) - |\w|),
\end{aligned}
\end{equation}
\begin{equation}\label{eq:Ru}
\begin{aligned}
  R_{\alpha,\,u\mathrm{-channel}} =&\frac{1}{2^7\pi^3E_k |\k| } \int_{m_{k'}}^\infty dE_{k'} \int_{\max(m_p,k'-k+m_{p'})}^\infty dE_p \int d|\v| \: S|M_u|^2 F\\
  &\times \theta(|\v| - \max(||\k|-|\p'||,||\k'|-|\p||)) \theta(\min(|\k|+|\p'|,|\k'|+|\p|)-|\v|) .
\end{aligned}
\end{equation}
After inserting the matrix element $S|M_x|^2$ and momentum distributions $F$, these reduced integrals are valid for any $2 \to 2$ process.


\subsection{The full collision terms}

The matrix elements for all elastic and inelastic processes involving $\nu_\alpha$ at temperatures $T  \lesssim m_\mu$ are  summarised in \tref{tab:matrixel}.
 These have been computed at various times by several different groups~\cite{Flaig:1989,Hannestad:1995rs,Dolgov:1997mb}, but can be easily obtained from first principles
in the four-fermion limit.  Using these matrix elements and the expressions~\eref{eq:Rs}, \eref{eq:Rt} and \eref{eq:Ru}, we can now determine the contribution of each process to the repopulation integral.  The results are as follows.
 
\begin{enumerate} 
\item $\nu_\alpha(k) \nu_\beta(p) \rightarrow \nu_\alpha(k') \nu_\beta(p')$:
\begin{equation}
\begin{aligned}
\label{eq:finalRsbeta}
&  R_{\alpha,s,\beta}  = \frac{ G_F^2 }{2(2\pi)^3E_k^2}\int_{0}^\infty dE_{k'} \int_{\max(0,E_{k'}-E_k)}^\infty dE_p \int_{\max(|E_k-E_p|,|2E_{k'}-E_{p}-E_k|)}^{E_k+E_p} d|\q|\\ 
&\hspace{3mm} \times \Big[(E_p+E_k)^2-|\q|^2 \Big]^2\\
& \hspace{7mm} \times  \Big[ f(E_{k'})f(E_{p'})(1-f(E_k))(1-f(E_p))-f(E_k)f(E_p)(1-f(E_{k'}))(1-f(E_{p'}) ) \Big]  .
\end{aligned}
\end{equation}

\item  $\nu_\alpha(k) \bar\nu_\beta(p) \rightarrow \nu_\alpha(k') \bar\nu_\beta(p')$:
\begin{equation}
\begin{aligned}
\label{eq:finalRsbetabar}
&  R_{\alpha,s,\bar\beta} =\frac{ G_F^2 }{2(2\pi)^3E_k^2}\int_{0}^\infty dE_{k'} \int_{\max(0,E_{k'}-E_k)}^\infty dE_p \int_{|E_{k'}-E_p|}^{\min(2E_k+E_{p}-E_{k'},E_{k'}+E_p)} d|\v|\\ 
&\hspace{3mm}\times \Big[(E_p-E_{k'})^2-|\v|^2 \Big]^2\\
&\hspace{7mm} \times \Big[f(E_{k'})f(E_{p'})(1-f(E_k))(1-f(E_p))-f(E_k)f(E_p)(1-f(E_{k'}))(1-f(E_{p'})) \Big] .
\end{aligned}
\end{equation}

\item $\nu_\alpha(k) \nu_\alpha(p) \rightarrow \nu_\alpha(k') \nu_\alpha(p')$:
\begin{equation}
\begin{aligned}
\label{eq:finalRsalpha}
 & R_{\alpha,s,\alpha} =\frac{ G_F^2 }{(2\pi)^3E_k^2}\int_{0}^\infty dE_{k'} \int_{\max(0,E_{k'}-E_k)}^\infty dE_p \int_{\max(|E_k-E_p|,|2E_{k'}-E_{p}-E_k|)}^{E_k+E_p} d|\q|\\ 
&\hspace{3mm} \times \Big[(E_p+E_k)^2-|\q|^2 \Big]^2\\
&\hspace{7mm} \times  \Big[ f(E_{k'})f(E_{p'})(1-f(E_k))(1-f(E_p))-f(E_k)f(E_p)(1-f(E_{k'}))(1-f(E_{p'})) \Big].
\end{aligned}
\end{equation}

\item $\nu_\alpha(k) \bar\nu_\alpha(p) \rightarrow \nu_\alpha(k') \bar\nu_\alpha(p')$:
\begin{equation}
\begin{aligned}
\label{eq:finalRsalphabar}
&  R_{\alpha,s,\bar\alpha} =\frac{ 2G_F^2 }{(2\pi)^3E_k^2}\int_{0}^\infty dE_{k'} \int_{\max(0,E_{k'}-E_k)}^\infty dE_p \int_{|E_{k'}-E_p|}^{\min(2E_k+E_{p}-E_{k'},E_{k'}+E_p)} d|\v| \\
&\hspace{3mm} \times \Big[ (E_p-E_{k'})^2-|\v|^2 \Big]^2\\
&\hspace{7mm} \times \Big[ f(E_{k'})f(E_{p'})(1-f(E_k))(1-f(E_p))-f(E_k)f(E_p)(1-f(E_{k'}))(1-f(E_{p'}) )\Big]  .
\end{aligned}
\end{equation}

\item $ \nu_\alpha(k) e^-(p) \rightarrow \nu_\alpha(k') e^-(p')$:
\begin{equation}
\begin{aligned}
\label{eq:finalRse}
&  R_{\alpha,s,e^-} =\frac{G_F^2}{2(2\pi)^3E_k^2}\int_{0}^\infty dE_{k'} \int_{\max(m_e,E_{k'}-E_k+m_{e})}^\infty dE_p\\
  & \times \bigg[ \int d|\q| \theta(|\q|-\max(|E_k-|\p||,|E_{k'}-|\p'||)) \theta(\min(E_k+|\p|,E_{k'}+|\p'|)-|\q|)\\ 
    &\hspace{10mm} \times (2x_W\pm 1)^2 ((E_p+E_k)^2-|\q|^2 - m_e^2)^2 \\
    &\hspace{7mm}+ \int d|\v| \theta(|\v|-\max(|E_k-|\p'||,E_{k'}-|\p||)) \theta(\min(E_k+|\p'|,E_{k'}+|\p|)-|\v|)\\ 
    &\hspace{10mm} \times 4 x_W^2 ((E_p-E_{k'})^2 -|\v|^2 -m_e^2)^2\\
    &\hspace{7mm}+ \int d|\w| \theta(|\w|-\max(||\p|-|\p'||,|E_k-E_{k'}|)) \theta(\min(|\p|+|\p'|,E_k+E_{k'})-|\w|) \\ 
    &\hspace{10mm} \times 4m_e^2(2x_W\pm 1)x_W ((E_k-E_{k'})^2 - |\w|^2) \bigg]\\
    &\times \Big[f(E_{k'})f(E_{p'})(1-f(E_k))(1-f(E_p)) - f(E_k)f(E_p)(1-f(E_{k'}))(1-f(E_{p'})) \Big] .
\end{aligned}
\end{equation}

\item $\nu_\alpha(k) e^+(p) \rightarrow \nu_\alpha(k') e^+(p')$:
\begin{equation}
\begin{aligned}
\label{eq:finalRsp}
&  R_{\alpha,s,e^+} =\frac{G_F^2}{2(2\pi)^3E_k^2}\int_{0}^\infty dE_{k'} \int_{\max(m_e,E_{k'}-E_k+m_{e})}^\infty dE_p\\
  &\times \bigg[ \int d|\q| \theta(|\q| - \max(|E_k-|\p||,|E_{k'}-|\p'||)) \theta(\min(E_k+|\p|,E_{k'}+|\p'|)-|\q|)\\ 
  &\hspace{10mm} \times 4 x_W^2 ((E_p+E_k)^2-|\q|^2 - m_e^2)^2\\
  &\hspace{7mm}+ \int d|\v| \theta(|\v|-\max(|E_k-|\p'||,E_{k'}-|\p||)) \theta(\min(E_k+|\p'|,E_{k'}+|\p|) - |\v|)\\
  &\hspace{10mm} \times (2x_W\pm 1)^2 ((E_p-E_{k'})^2 -|\v|^2 -m_e^2)^2\\
  &\hspace{7mm}+ \int d|\w| \theta(|\w| - \max(||\p|-|\p'||,|E_k-E_{k'}|)) \theta(\min(|\p|+|\p'|,E_k+E_{k'})- |\w|)\\
  &\hspace{10mm} \times 4m_e^2(2x_W\pm 1)x_W((E_k-E_{k'})^2 - |\w|^2) \bigg]\\
 &\times \Big[ f(E_{k'})f(E_{p'})(1-f(E_k))(1-f(E_p)) - f(E_k)f(E_p)(1-f(E_{k'}))(1-f(E_{p'})) \Big].
\end{aligned}
\end{equation}

\item $\nu_\alpha(k) \bar\nu_\alpha(p) \rightarrow \nu_\beta(k') \bar\nu_\beta(p')$:
\begin{equation}
\begin{aligned}
\label{eq:finalRabeta}
 & R_{\alpha,a,\beta} =\frac{ G_F^2 }{2(2\pi)^3E_k^2}\int_{0}^\infty dE_{p} \int_{0}^{E_k+E_p} dE_{k'} \int_{|E_{k'}-E_p|}^{\min(2E_k+E_{p}-E_{k'},E_{k'}+E_p)} d|\v|\\
 &\hspace{3mm} \times \Big[(E_p-E_{k'})^2-|\v|^2 \Big]^2\\ 
 &\hspace{7mm} \times \Big[f(E_{k'})f(E_{p'})(1-f(E_k))(1-f(E_p))-f(E_k)f(E_p)(1-f(E_{k'}))(1-f(E_{p'})) \Big] .
\end{aligned}
\end{equation}

\item $\nu_\alpha(k) \bar\nu_\alpha(p) \rightarrow e^-(k') e^+(p')$:
\begin{equation}
\begin{aligned}
\label{eq:finalRae}
&  R_{\alpha,a,e} =\frac{G_F^2}{2(2\pi)^3E_k^2}\int_{\min(0,2m_e-E_k)}^\infty dE_{p} \int_{m_e}^{E_k+E_p-m_e} dE_{k'}\\
  &\times \bigg[ \int d|\q| \theta(|\q|- \max(|E_k-E_p|,||\k'|-|\p'||)) \theta(\min(E_k+E_p,|\k'|+|\p'|)- |\q|) \\ 
    & \hspace{10mm}\times(2x_W \pm 1)4x_Wm_e^2 (|\q|^2 -(E_p+E_k)^2) \\
    &\hspace{7mm} + \int d|\v| \theta(|\v|-\max(|E_k-|\p'||,||\k'|-E_p|)) \theta(\min(E_k+|\p'|,|\k'|+E_p)- |\v|) \\
    &\hspace{10mm} \times 4x_W^2 ((E_p-E_{k'})^2-|\v|^2 -m_e^2)^2\\
    &\hspace{7mm}+ \int d|\w| \theta(|\w|- \max(|E_p-|\p'||,|E_k-|\k'||)) \theta(\min(E_p+|\p'|,E_k+|\k'|) - |\w|)\\
    &\hspace{10mm} \times (2x_W\pm 1)^2 ((E_k-E_{k'})^2 - |\w|^2 - m_e^2)^2 \bigg]\\
    &\times  \Big[f(E_{k'})f(E_{p'})(1-f(E_k))(1-f(E_p))  - f(E_k)f(E_p)(1-f(E_{k'}))(1-f(E_{p'})) \Big].
\end{aligned}
\end{equation}

\end{enumerate}
We remind the reader again that $|\p'|$ is not a free parameter, but is constrained by energy conservation.

The integrals over $|\q|$, $|\v|$ and $|\w|$ can be evaluated analytically, and it turns out that they fall into two different functional forms:
\begin{align*}
  \int dx (a-x^2)^2 &=  a^2 x - \frac{2}{3} a x^3 + \frac{x^5}{5} + \textrm{constant} ,\\
  \int dx (a-x^2) &= a x - \frac{x^3}{3} + \textrm{constant} .
\end{align*}
The remaining two integrals over $E_p$ and $E_{k'}$ must be performed numerically, although as discussed in \sref{sec:numimp} judicious assumptions about certain distribution functions in the integrand make it possible to pre-evaluate some of the integrals only once, rather than evaluating them in real time simultaneously with the numerical solution of the QKEs.


\section{Repopulation and damping coefficients in the A/S approximation}
\label{sec:dampingcoeff}

We summarise here the full expressions for the dimensionless repopulation and damping coefficients in the A/S approximation discussed in section \ref{sec:approx_intro}.
The quantity $A \equiv 2\pi/\int d \Pi_k k\: f_0(k)$ is a normalisation factor.
\begin{align*}
  C_{\alpha,a} &= A
  \int d \Pi_k d \Pi_{k'} d \Pi_{p'} d \Pi_p \; \delta_E(kp|k'p') 
  \sum_{i} \mathcal{V}^2[\nu_\alpha(k),\bar\nu_\alpha(p)| i(k'),\bar i(p')] f_0(p) f_0(k) ,\\
  C_{\alpha,s} &= A
  \int d \Pi_k d \Pi_{k'} d \Pi_{p'} d \Pi_p \; \delta_E(kp|k'p')
  \sum_{j \neq \nu_\alpha, \bar{\nu}_\alpha} \mathcal{V}^2[\nu_\alpha(k),j(p)|\nu_\alpha(k'),j(p')] f_0(p) f_0(k) , \\
  C_{\alpha, \nu} &= A
  \int d \Pi_k d \Pi_{k'} d \Pi_{p'}d \Pi_p \; \delta_E(kp|k'p')
  \sum_{j = \nu_\alpha, \bar{\nu}_\alpha} \mathcal{V}^2[\nu_\alpha(k),j(p)|\nu_\alpha(k'),j(p')] f_0(p) f_0(k) ,
 \end{align*}
  \begin{align*}
  C_{\alpha,0} &= A
  \int d \Pi_k d \Pi_{k'}d \Pi_{p'}d \Pi_p \; \delta_E(kp|k'p') f_0(k) \nonumber\\
&\hspace{2.8cm}   \times \Big[ \sum_{i} \mathcal{V}^2[\nu_\alpha(k),\bar\nu_\alpha(p)| i(k'),\bar i(p')] f_0(k') f_0(p')  \nonumber\\
  &\hspace{3.5cm} 
+ \sum_{j \neq \nu_\alpha, \bar{\nu}_\alpha} \mathcal{V}^2[\nu_\alpha(k),j(p)|\nu_\alpha(k'),j(p')] f_0(p)(1-f_0(p')) \Big], \\
  C_{\alpha,1} &=A
  \int d \Pi_k d \Pi_{k'} d \Pi_{p'} d \Pi_p \; \delta_E(kp|k'p') f_0(k) \nonumber\\
  &\hspace{2.8cm} \times \Big[ \sum_{i} \mathcal{V}^2[\nu_\alpha(k),\bar\nu_\alpha(p)| i(k'),\bar i(p')] f_0(p) (1-f_0(p') -f_0(k'))
  \nonumber\\
  &\hspace{3.5cm} + \sum_{j \neq \nu_\alpha, \bar{\nu}_\alpha} \mathcal{V}^2[\nu_\alpha(k),j(p)|\nu_\alpha(k'),j(p')] f_0(k') (f_0(p')-f_0(p)) \nonumber\\
  & \hspace{3.5cm}  + \sum_{j = \nu_\alpha, \bar{\nu}_\alpha} \mathcal{V}^2[\nu_\alpha(k),j(p)|\nu_\alpha(k'),j(p')]  f_0(p) \Big],\\
  C_{\alpha,2} &=A
  \int d \Pi_k d \Pi_{k'}d \Pi_{p'}d \Pi_p \; \delta_E(kp|k'p') f_0(k) \nonumber\\
  &\hspace{1.3cm} \times \sum_{j = \nu_\alpha, \bar{\nu}_\alpha} \mathcal{V}^2[\nu_\alpha(k),j(p)|\nu_\alpha(k'),j(p')] (f_0(k') f_0(p')- f_0(k') f_0(p) - f_0(p) f_0(p') ).
\end{align*}
See equations~\eref{eq:Ras} and \eref{eq:Das} for the implementation of these coefficients in the A/S repopulation and damping terms.


\bibliography{collision}

\end{document}